\documentclass[mnsc, nonblindrev]{informs3} % current default for manuscript submission

\OneAndAHalfSpacedXI
\usepackage{natbib}
\bibpunct[, ]{(}{)}{,}{a}{}{,}%
\def\bibfont{\small}%
\def\BIBand{and}%

\TheoremsNumberedThrough     % Preferred (Theorem 1, Lemma 1, Theorem 2)
\ECRepeatTheorems

\EquationsNumberedThrough    % Default: (1), (2), ...
\usepackage{amsmath,amssymb,amsfonts}
\usepackage{mathtools}
\usepackage{algorithmic}
\usepackage{algorithm}
\usepackage{textcomp}
\usepackage{xcolor}	
\definecolor{DarkBlue}{rgb}{0,0.08,0.45}
\usepackage{multirow}
\usepackage{makecell}
\usepackage{tabularx}
\usepackage{latexsym}
\usepackage{subcaption}
\usepackage{comment}

\usepackage{bbm}
\usepackage{booktabs}
\PassOptionsToPackage{hyphens}{url}
\usepackage[backref = false, bookmarks, colorlinks = true, plainpages = false, citecolor = DarkBlue, urlcolor = DarkBlue, filecolor = DarkBlue, linkcolor = DarkBlue, breaklinks]{hyperref}

\def\BibTeX{{\rm B\kern-.05em{\sc i\kern-.025em b}\kern-.08em
    T\kern-.1667em\lower.7ex\hbox{E}\kern-.125emX}}

\newlength\myindent
\usepackage[T1]{fontenc}
\usepackage{lmodern}
\usepackage[utf8]{inputenc} % allow utf-8 input
\usepackage[T1]{fontenc}    % use 8-bit T1 fonts
\usepackage{hyperref}       % hyperlinks
\usepackage{url}            % simple URL typesetting
\usepackage{booktabs}       % professional-quality tables
\usepackage{amsfonts}       % blackboard math symbols
\usepackage{nicefrac}       % compact symbols for 1/2, etc.
\usepackage{microtype}      % microtypography
\usepackage{graphicx}
\usepackage{amsmath}
\usepackage{rotating}
\usepackage{mathrsfs}
\usepackage{array,multirow}
\usepackage[justification=centering]{caption}
\usepackage{bbm}

\usepackage{makecell}

\usepackage{tabularx}

\usepackage{mathtools}
\DeclarePairedDelimiterX{\norm}[1]{\lVert}{\rVert}{#1}

\usepackage{circuitikz}

\begin{document}
\TITLE{{\Large Diversity-Fair Online Selection}}
\ARTICLEAUTHORS{
 \AUTHOR{Ming Hu}
 \AFF{Rotman School of Management, University of Toronto, \EMAIL{ming.hu@rotman.utoronto.ca}}
\AUTHOR{Yanzhi Li}
\AFF{College of Business, City University of Hong Kong, \EMAIL{yanzhili@cityu.edu.hk}}
\AUTHOR{Tongwen Wu}
\AFF{Rotman School of Management, University of Toronto, \EMAIL{tw.wu@rotman.utoronto.ca}} 
 }

\ABSTRACT{ 
Online selection problems arise in applications such as crowdsourcing and recruitment, where decision makers may seek representation across multiple, potentially overlapping demographic or skill dimensions. We study diversity-fair online selection under adversarial arrivals. A recruiter must immediately and irrevocably decide whether to accept each candidate while selecting at most \(K\) candidates. Before arrivals begin, the recruiter observes aggregate marginal information: the total number of candidates contributing to each of the \(d\) diversity dimensions. When the candidate pool is large, this information may be estimated from demographic statistics of the applicant population. We evaluate the expected utilities across dimensions using the generalized mean \( M_p=(d^{-1}\sum_{k=1}^d U_k^p)^{1/p},  -\infty\le p\le 1, \) where \(U_k\) denotes the expected utility of dimension \(k\). We first study max-min fairness, corresponding to \(p=-\infty\). We prove that no online policy can achieve a competitive ratio better than \(O(1/\sqrt d)\) and develop a policy with a competitive ratio \(1/[4(2+\sqrt2)\sqrt d]\), establishing the optimal dependence on \(d\) up to a constant factor. Without exact marginal information, the optimal worst-case rate falls to \(\Theta(1/d)\), demonstrating the value of this information. Our policy first constructs an online fractional solution and then implements it using a standard online rounding scheme that preserves each dimension’s expected utility. We also extend the max-min analysis to nonbinary attributes and characterize the optimal dependence on their value range. Finally, we study generalized-mean objectives. For \(0\le p\le1\), we establish an optimal competitive ratio of \(\Theta(1/\log d)\). For each fixed finite negative mean \(p=-q\), where \(q>0\), our policy achieves \(d^{-q/(2q+1)}\) up to polylogarithmic factors, matching the exponent of the corresponding impossibility bound. Together, these results characterize how the attainable diversity guarantee varies with the fairness objective and quantify the value of aggregate marginal information under adversarial arrivals.

}
%	\begin{ABSTRACT}
%		Your abstract.
%	\end{ABSTRACT}

\maketitle 

%\vspace{-1em}

\section{Introduction}
Online selection problems arise in applications such as crowdsourcing and recruitment, where a decision maker must accept or reject candidates as they arrive. Classical models typically rank candidates by a scalar value and seek to select the best candidate or a subset of the highest-value candidates \citep{babaioff2008online}. In many applications, however, selection quality also depends on representation across multiple, potentially overlapping dimensions.

Consider assembling a focus group to evaluate a pilot television episode or recruiting participants for a research survey. A useful panel should represent different demographic groups, experiences, and perspectives so that its feedback reflects the target population. Because one participant may contribute to several dimensions simultaneously, the decision maker must account not only for the quality of individual candidates but also for the diversity of the selected group.

In these settings, the decision maker faces a fixed headcount capacity and must make immediate and irrevocable selection decisions. Although future candidate profiles are unknown, aggregate information about individual dimensions may be available from population statistics or the platform's candidate pool. Such information may reveal, for example, the total number of candidates possessing each demographic or skill attribute without revealing which attributes occur together or when the candidates will arrive. This leads to our central question: \emph{Given aggregate marginal information, what diversity guarantee is achievable under adversarial arrivals?}

To this end, we consider a setting in which candidates arrive one at a time over an unknown horizon of \(n\) arrivals. The decision maker (DM) may select at most \(K\) candidates to generate utility across \(d\) diversity dimensions. Candidate \(i\) arrives with a binary profile \(\mathbf{t}_i\in\{0,1\}^d\), where \(t_{ik}=1\) indicates that the candidate contributes to dimension \(k\). Because a candidate may contribute to multiple dimensions, several components of \(\mathbf t_i\) may equal one simultaneously. Upon observing the candidate's profile, the DM must immediately and irrevocably decide, possibly at random, whether to select the candidate. The arrival sequence is adversarial, and the capacity constraint must hold in every realization.

Let \(U_k\) denote the expected utility of dimension \(k\), equal to \(c_k\) times the expected number of selected candidates contributing to that dimension. For \(p\ne0\), we evaluate the utility vector \(\mathbf U=(U_1,\ldots,U_d)\) using the generalized mean \(M_p(\mathbf U)=(d^{-1}\sum_{k=1}^d U_k^p)^{1/p}\). The limiting cases are \(M_0(\mathbf U)=(\prod_{k=1}^d U_k)^{1/d}\) and \(M_{-\infty}(\mathbf U)=\min_{k\in[d]}U_k\). We first study max-min fairness, corresponding to \(p=-\infty\), and then consider every finite \(p\le1\).

We compare an online policy with an optimal offline benchmark that knows the complete arrival sequence in advance. For a fixed number of dimensions \(d\), the competitive ratio of an online policy is the infimum, over all instances with \(d\) dimensions, of the ratio between the policy's objective value and the offline optimal value.
Our focus is the \textit{advice-augmented scenario}. In addition to \(d\), \(K\), and the dimension coefficients \(\{c_k\}_{k\in[d]}\), the DM knows before the horizon the exact aggregate marginal count \(\sum_{i\in[n]}t_{ik}\) for each dimension \(k\). These counts reveal how frequently each attribute appears over the horizon, but not which attributes occur together within a candidate or when those candidates arrive. When the candidate pool is large, we can estimate this information from demographic statistics of the applicant population; we also quantify the cost of misspecified information. In the literature on online algorithms, such instance-dependent advance information is often called advice or a prediction \citep{jin2022online,balseiro2023single,banerjee2022online,banerjee2022proportionally}.
Under this information structure, we seek to characterize how the attainable competitive ratio varies with the generalized-mean parameter \(p\) and the number of dimensions \(d\).

\subsubsection*{Results.} 
Our work lies at the intersection of fairness-oriented allocation and online algorithms with predictions \citep{banerjee2022proportionally,banerjee2023online,barman2022universal,huang2025long}. The closest work is \citet{banerjee2022proportionally}, who study online public-goods allocation under proportional fairness, an objective closely related to the geometric mean \(M_0\), and establish tight competitive guarantees. Like public-goods allocation, selecting one candidate in our model can benefit multiple dimensions simultaneously.\footnote{Their model, however, permits fractional allocations and provides the algorithm with mildly different information about future arrivals. But these are not the fundamental differences.} However, we instead study online selection across the full range of generalized means, \(-\infty\le p\le1\), including the most egalitarian endpoint, max-min fairness. As \(p\) decreases below zero, the objective places progressively greater weight on dimensions with low utility, producing qualitatively different competitive guarantees and requiring new algorithmic ideas.

We organize our analysis around two regimes. We first establish matching-order bounds for max-min fairness, examine the value and robustness of aggregate marginal advice, and extend the model to nonbinary attributes. We then study finite generalized means and characterize how the attainable competitive ratio varies with \(p\).

\paragraph{Max-min fairness.}

To assess the value of aggregate marginal advice for diversity-fair online selection, we first show that, under max-min fairness, without exact advice, no policy can achieve a competitive ratio better than $O(1/d)$, matching the trivial $1/d$ guarantee up to a constant factor. This raises a central question for this multidimensional objective. How much can exact marginal information improve the competitive guarantee? We answer this question by developing a policy that achieves a competitive ratio of
\(
    1/ (4(2+\sqrt2)\sqrt d).
\)
Complementing this guarantee, we show that no policy can achieve a competitive ratio better than $O(1/\sqrt d)$, even with exact marginal advice. Thus, our guarantee is tight up to a constant factor and characterizes the value of marginal information for this fairness-driven online decision-making.

We also extend the binary-attribute model to nonbinary attributes with \(t_{ik}\in\{0\}\cup[\ell,u]\). In this setting, the DM receives exact weighted marginal advice \(S_k=\sum_{i\in[n]}t_{ik}\) for each dimension \(k\). Letting \(\kappa=u/\ell\), our modified policy achieves a competitive ratio of \(1/[4(2+\sqrt2)\sqrt{d\kappa}]\), while an impossibility result shows that no policy can achieve a competitive ratio better than \(O(1/\sqrt{d\min\{\kappa,d\}})\). For \(1\le\kappa\le d\), these bounds match up to a constant factor, establishing the tight dependence \(\Theta(1/\sqrt{d\kappa})\).

\paragraph{Generalized means.}
We next return to the binary-attribute model with exact aggregate marginal advice and extend the analysis from max-min fairness to finite generalized means. Table~\ref{tab:intro_generalized_mean} summarizes our policy guarantees and the corresponding impossibility bounds across the full range \(-\infty\le p\le1\), including max-min fairness for comparison. For finite negative means, we write \(p=-q\) with \(q>0\).\footnote{The notation of \(O_q(\cdot)\) and \(\Omega_q(\cdot)\) suppresses multiplicative constants that may depend on \(q\) but not on \(d\).}

\begin{table}
\centering
\caption{Competitive ratios for generalized-mean objectives.}
\label{tab:intro_generalized_mean}
\small
\begin{tabularx}{\textwidth}{@{}lXcc@{}}
\toprule
Parameter & Diversity objective & Policy guarantee & Impossibility bound \\
\midrule
$0<p\le1$
& Positive generalized mean
& \makecell[c]{$\displaystyle \frac{1}{1+4\log(2d+2)}$}
& \makecell[c]{$O(1/\log d)$} \\
\\
$p=0$
& Geometric mean (Nash social welfare)
& \makecell[c]{$\displaystyle \frac{1}{4\log(2d+2)}$}
& \makecell[c]{$O(1/\log d)$} \\
\\
$p=-q$, $q>0$
& Finite negative generalized mean
& \makecell[c]{$\displaystyle \frac{\Omega_q(1)}{d^{q/(2q+1)}\mathrm{polylog}(d)}$}
& \makecell[c]{$O_q\!\left(d^{-q/(2q+1)}\right)$} \\
\\
$p=-\infty$
& Max-min fairness
& \makecell[c]{$\displaystyle \frac{1}{4(2+\sqrt2)\sqrt d}$}
& \makecell[c]{$O(1/\sqrt d)$} \\
\bottomrule
\end{tabularx}
\end{table}

Together, these results reveal how the attainable competitive ratio changes with the generalized-mean parameter \(p\). As \(p\) decreases from \(1\) to \(0\), the optimal rate remains \(\Theta(1/\log d)\). At \(p=1\), the objective is the arithmetic mean and hence linear; the problem resembles classical single-leg revenue management, where the same logarithmic barrier arises \citep{ball2009toward}. At \(p=0\), the objective becomes the geometric mean, and the logarithmic rate matches that established in the public-goods setting by \citet{banerjee2022proportionally}.\footnote{Their model permits fractional allocation of each arriving public good and differs from ours in the information available to the algorithm. We adapt their arguments to our online selection setting in Theorem~\ref{thm:geometric_mean}.}

Once \(p\) decreases below zero, the dependence on \(d\) changes from logarithmic to polynomial. Writing \(p=-q\) with \(q>0\), our policy achieves \(d^{-q/(2q+1)}\) up to polylogarithmic factors, matching the exponent of the corresponding impossibility bound. As \(p\) decreases further, equivalently as \(q\) increases, the polynomial exponent \(q/(2q+1)\) approaches \(1/2\). The attainable rate therefore approaches the polynomial rate of \(d^{-1/2}\) for max-min fairness at \(p=-\infty\).

\subsubsection*{Technical contributions.}
Because our objective is defined by expected dimension utilities, online level-set rounding \citep{srinivasan2023online} allows us to focus on designing an online fractional algorithm. This connects our problem to the online public-goods allocation model of \citet{banerjee2022proportionally}, which can be viewed as studying such online fractional algorithms under a geometric-mean objective.

Our main technical contribution is an order-optimal policy under max-min fairness. 
The set-aside greedy policy of \citet{banerjee2022proportionally} allocates half of the budget uniformly across arriving public goods and reserves the other half for greedy allocations made only when the marginal gain exceeds a static threshold.
In contrast, because the max-min objective is determined by the lowest-utility dimension, we use a guess-then-optimize framework, formulated as a principal-agent design, in which the principal creates one agent for each geometric guess of the offline optimum and assigns capacity to that agent. For each fixed guess, the corresponding agent divides its budget between \textit{controlled greedy} and \textit{minimalist protection}. Controlled greedy allocates only when the current candidate contributes to sufficiently many dimensions below the target, thereby obtaining broad coverage from each unit of capacity. Minimalist protection operates dimension by dimension. It uses the aggregate advice to track each dimension's realized utility and remaining opportunities and allocates only when its target would otherwise become unattainable. The two components are complementary. Controlled greedy exploits candidates who benefit many dimensions simultaneously, whereas minimalist protection safeguards the few dimensions that remain underrepresented.

Principal-agent designs are common in online algorithms, but running multiple guesses in parallel typically incurs a loss proportional to the number of guesses. Our policy avoids this logarithmic loss through a nonuniform budget allocation. The principal assigns each agent a budget proportional to the square root of its guess and calibrates the controlled-greedy threshold to that budget. These budgets form a geometric series within the total capacity, and the principal combines the agents' allocations through a capped sum. Consequently, the agent with the successful guess retains its full $\Theta(1/\sqrt d)$ guarantee without an additional averaging loss.

For $0\le p\le1$, we extend the set-aside greedy framework of \citet{banerjee2022proportionally} by calibrating the set-aside allocation with marginal advice and, for $0<p\le1$, replacing the static threshold with a dynamic threshold scaled by the current $p$-mean potential. For finite negative means, minimalist protection reappears within a more elaborate principal-agent design involving global and group-specific guesses. Each agent runs a Nashian allocation on a reference set and then applies minimalist protection only to dimensions that remain below a prescribed floor.

\paragraph{Organization of the paper.}
Section~\ref{sec:literature} reviews the related literature. Section~\ref{sec:model} introduces the online selection model, the offline benchmark, the role of marginal advice, and the online-rounding reduction. Section~\ref{sec:general} develops and analyzes our policy for max-min fairness and extends it to nonbinary attributes. Section~\ref{sec:generalized_mean} studies nonnegative and negative generalized-mean objectives. Section~\ref{sec:conclusion} concludes. The electronic companion contains omitted proofs and additional algorithmic details.

\section{Literature Review}
\label{sec:literature}

\subsubsection*{Fairness and diversity in selection and allocation.}
Fair allocation is commonly formulated through max-min or proportional objectives, which capture different trade-offs between individual utilities and aggregate efficiency \citep{bertsimas2011price}. Alternatively, some operational models optimize efficiency subject to fairness constraints, as in assortment planning \citep{chen2022fair}. In diversity-aware selection, fairness is often imposed through representation requirements. Citizens' assemblies use demographic quotas \citep{flanigan2021fair}, school admissions use affirmative-action constraints \citep{bonet2024explainable}, and hiring and search models impose interview-stage representation or demographic-parity constraints \citep{Farajol2025rooney,aminian2023markovian}. In contrast to these constraint-based diversity models, we adopt an objective-based formulation that optimizes the generalized mean of dimension utilities.

Online fairness has been studied across several operational settings. Dynamic rationing and resource-division models examine how limited resources should be shared as demand or the population evolves \citep{lien2014sequential,manshadi2023fair,sinclair2022sequential,banerjee2023online,vardi2022dynamic}. Regularized online allocation incorporates fairness directly into the objective \citep{balseiro2021regularized}, while other work considers fairness in online matching, ride-hailing, and the allocation of indivisible goods \citep{freund2023group,ma2023fairness,nanda2020balancing,benade2024fair}. Our setting is distinguished by adversarial accept-or-reject decisions, overlapping diversity dimensions, and advice limited to the aggregate total for each dimension. In particular, selecting one candidate may simultaneously increase utilities across several dimensions.

\subsubsection*{Online algorithms with advice.}
Our advice-augmented setting belongs to the growing literature that uses predictions to improve online decisions under otherwise adversarial inputs \citep{mahdian2012online,bamas2020primal,jin2022online,balseiro2023single,feng2024robust}. Predictions may summarize future demand, an offline solution, or another instance-dependent quantity. The resulting guarantees quantify how an algorithm benefits from accurate information and, in some models, how its performance degrades when that information is inaccurate.

The online public-goods allocation model of \citet{banerjee2022proportionally} is structurally closest to ours. In that model, a decision maker allocates a limited budget among sequentially arriving public goods, and one investment can simultaneously increase the utilities of several agents. Similarly, selecting one candidate in our model consumes shared capacity while benefiting every dimension represented by that candidate. Both models also provide the algorithm with aggregate information about individual utilities or dimensions without revealing the complete future instance. Their model, however, permits fractional investment and focuses on proportional fairness, whereas ours makes discrete selections under a hard cardinality constraint and studies generalized means over \(-\infty\le p\le1\). After our online rounding reduction, the fractional formulations are closely related, but the max-min and negative-mean objectives require different ways of protecting low-utility dimensions.

Another related problem is the online fair allocation of divisible private goods. Goods arrive sequentially, and each good must be divided among agents. Unlike public goods, a fraction assigned to one agent cannot also benefit another, so agents compete for each arriving resource. \citet{banerjee2022online} maximize Nash social welfare using predictions of agents' aggregate utilities, while \citet{barman2022universal} and \citet{huang2025long} study generalized-mean welfare over the broader range \(-\infty\le p\le1\). These works use forms of advice and fairness objectives similar to ours, but the underlying allocation structure differs: in our setting, selecting a candidate can benefit multiple dimensions simultaneously, rather than requiring the dimensions to compete for mutually exclusive shares.

Our policies share several high-level ideas with the private-goods literature. \citet{barman2022universal} combine a guess-then-optimize framework, greedy allocation, and targeted allocations to vulnerable agents, paralleling our use of geometric guesses and minimalist protection. \citet{huang2025long} combine Nashian allocation with auxiliary allocations for low-utility agents, a template related to our negative-mean policy. Our shared selection capacity and overlapping dimension benefits, however, require different methods for budgeting and combining these components. The models also exhibit different competitive-ratio spectra. For fixed \(p>0\), divisible private-goods allocation admits a constant competitive ratio, whereas our problem has a \(\Theta(1/\log d)\) ratio. For \(p=-q<0\), the private-goods ratio scales as \(d^{-q/(q+1)}\) for \(0<q\le1\) and as \(d^{-1/2}\) for \(q\ge1\), compared with \(d^{-q/(2q+1)}\) in our problem. The dependence on \(d\) coincides at \(p=0\) and \(p=-\infty\) but differs throughout other values of \(p\).

Finally, \citet{cohen2023general} obtain near-optimal guarantees for a broad class of objectives using one learned proportional-allocation parameter per agent. Their predictions are tailored to a specific allocation rule, whereas ours directly specify marginal counts across the arriving candidates.

% Given similar forms of advice \citep{banerjee2022online, barman2022universal,huang2025long}, our online selection mainly differs from the private-goods allocation in the allocation structure: selecting a candidate uses one unit of the common headcount capacity while simultaneously increasing the utilities of all dimensions represented by that candidate. 

\subsubsection*{Online selection and related algorithmic techniques.}
Classical online-selection models include the secretary problem and its generalizations, which assume a random arrival order \citep{ferguson1989solved,babaioff2008online}; random-order online packing \citep{kesselheim2014primal}; and prophet inequalities, which assume known distributions and compare an online algorithm with a clairvoyant benchmark \citep{krengel1978semiamarts,samuel1984comparison,ma2024randomized}. Information about future arrivals, whether known in advance or learned from observations under a random arrival order, can be particularly valuable for fairness objectives. For example, \citet{ma2023fairness} study online matching under known i.i.d.\ arrivals and maximize the minimum expected matching rate across possibly overlapping groups, obtaining a constant competitive ratio. Our model instead permits an adversarial arrival order; the only information about future arrivals available to the DM consists of aggregate dimension totals, not the distribution of candidate types. Consequently, even exact marginal advice does not yield a dimension-independent guarantee.

Under adversarial arrivals, online bipartite matching serves as a canonical model \citep{mehta2013online,huang2024online}. Related work develops online rounding methods that convert fractional decisions into feasible randomized allocations, including online correlated selection \citep[Section~5.4]{huang2024online}, online contention-resolution schemes \citep{feldman2021online}, and online dependent rounding \citep{srinivasan2023online}. Because our capacity constraint is a uniform matroid, the online level-set rounding scheme of \citet{srinivasan2023online} preserves the fractional selection marginals while satisfying the headcount capacity. We therefore use online rounding as a standard reduction and focus our core technical analysis on designing the fractional policy.

\section{Model and Preliminaries}
\label{sec:model}
This section formulates the diversity-fair online selection problem, defines the offline benchmark and competitive ratio, and introduces marginal advice. We then establish impossibility results that characterize both the value and the limitations of this advice. Finally, we use standard online rounding to reduce the randomized selection problem to constructing a feasible online fractional solution.

\subsection{Problem Formulation}
\label{subsec:preli}
\subsubsection{Online selection process.}
Let $d,n\in\mathbb{N}^+$ denote the number of diversity dimensions and the number of candidate arrivals, respectively, and write $[m]\triangleq\{1,\ldots,m\}$ for any $m\in\mathbb{N}^+$. Before the arrival process begins, the decision maker (DM) knows the number of dimensions $d$, a positive integer headcount capacity $K$, and positive utility coefficients $\{c_k\}_{k\in[d]}$. The horizon length $n$ and the future candidate profiles are unknown. Section~\ref{subsec:population} specifies the additional information available to the DM.

Candidates arrive sequentially in an adversarial order. We consider an oblivious adversary that fixes the horizon length and the entire arrival sequence before the process begins. When candidate $i\in[n]$ arrives, the DM observes the binary profile $\mathbf{t}_i=(t_{i1},\ldots,t_{id})\in\{0,1\}^d$, where $t_{ik}=1$ means that candidate $i$ contributes to dimension $k$. A candidate may contribute to multiple dimensions. After observing $\mathbf{t}_i$, the DM must immediately and irrevocably decide, possibly at random, whether to select the candidate. If $\mathcal A$ denotes the resulting randomly selected set, feasibility requires $|\mathcal A|\le K$.

Because multiplying all utility coefficients by the same positive constant does not affect any competitive ratio, we normalize $\min_{k\in[d]}c_k=1$. An instance is therefore represented by 
\(
    \mathcal I=
    \bigl(d,n,K,\{c_k\}_{k\in[d]},\{\mathbf t_i\}_{i\in[n]}\bigr).
\)

\subsubsection{Diversity objective.}
\label{subsubsec:diversity}

We first evaluate diversity using max-min fairness, which emphasizes the least-served dimension. For any candidate set $\mathcal X\subseteq[n]$, define its coverage of dimension $k$ as
\[
    \phi_k(\mathcal X)
    \triangleq
    \sum_{i\in\mathcal X}t_{ik}.
\]
Let $\mathcal A$ denote the random set selected by an online policy. The expected utility of dimension $k$ is
\(
    c_k\mathbb E[\phi_k(\mathcal A)],
\)
where the expectation is taken over the policy's internal randomization. The policy's performance on instance $\mathcal I$ is then
\begin{align}
\label{eq:dfn_alg}
    \texttt{ALG}(\mathcal I)
    \triangleq
    \min_{k\in[d]}c_k\mathbb E[\phi_k(\mathcal A)].
\end{align}
Section~\ref{sec:generalized_mean} replaces the minimum in \eqref{eq:dfn_alg} with a generalized mean of the same expected dimension-utility vector, thereby encompassing objectives such as Nash social welfare and utilitarian welfare.

\begin{remark}
The objective in \eqref{eq:dfn_alg} measures \emph{ex ante} fairness: each dimension is evaluated by its expected utility before the policy's random choices are realized. A stronger, ex post-oriented alternative would place the minimum inside the expectation,
\(
    \mathbb E\!\left[
        \min_{k\in[d]}c_k\phi_k(\mathcal A)
    \right].
\)
For every policy, this alternative is no larger than $\min_k c_k\mathbb E[\phi_k(\mathcal A)]$ and requires the dimensions to be balanced within the same realized selected set. We focus on ex ante fairness because it admits an exact fractional characterization of the randomized offline benchmark, as established in Lemma~\ref{lemma:equivalence}. The computational difficulty of deterministic and ex post formulations is discussed in Appendix~\ref{subsec:intract}.
\end{remark}

\subsubsection{Offline benchmark.}

We evaluate an online policy against an optimal offline policy that knows the entire instance $\mathcal I$ in advance. The offline policy may randomize, but its selected set must satisfy the same hard capacity constraint in every realization. Let $\texttt{OFF}(\mathcal I)$ denote the offline optimum.

Because the objective depends only on the marginal selection probabilities, the offline benchmark admits an exact fractional formulation. For any vector $\mathbf x\in[0,1]^n$, define
\[
    \texttt{LU}_{\mathcal I}(\mathbf x)
    \triangleq
    \min_{k\in[d]}
    c_k\sum_{i\in[n]}x_it_{ik}.
\]
We then consider
\begin{align}
\tag{Fluid}\label{eq:fluid}
\texttt{OPT}(\mathcal I)
\triangleq
\max_{\mathbf x}\quad
& \texttt{LU}_{\mathcal I}(\mathbf x)\\
\text{s.t.}\quad
& \sum_{i=1}^n x_i\le K,
\label{eq:capacity_constraint}\\
& 0\le x_i\le1,
\qquad i\in[n].
\label{eq:prob_candidate}
\end{align}
Here, $x_i$ represents the marginal probability of selecting candidate $i$. Let $\mathbf x^*$ denote an optimal solution to \eqref{eq:fluid}, so that
\[
    \texttt{OPT}(\mathcal I)
    =
    \texttt{LU}_{\mathcal I}(\mathbf x^*).
\]

\begin{lemma}[\sc Offline-Benchmark Equivalence]
\label{lemma:equivalence}
For every instance $\mathcal I$,
\(
    \emph{\texttt{OFF}}(\mathcal I)
    =
    \emph{\texttt{OPT}}(\mathcal I).
\)
\end{lemma}

To prove Lemma~\ref{lemma:equivalence}, observe first that the selection marginals of any feasible offline policy form a feasible solution to \eqref{eq:fluid} with the same objective value. Conversely, marginal-preserving dependent rounding implements any feasible fractional solution as a random set of size at most $K$ \citep{gandhi2006dependent}. The complete proof is provided in Appendix~\ref{subsec:proof_equiv}.

\subsubsection{Competitive ratio.}
\label{subsubsec:cr}
For each $d\in\mathbb N^+$, let $\mathscr I_d$ denote the set of instances with $d$ diversity dimensions. The competitive ratio of an online policy $\texttt{ALG}$ is
\[
    \operatorname{CR}_{\texttt{ALG}}(d)
    \triangleq
    \inf_{\substack{\mathcal I\in\mathscr I_d\\
                    \texttt{OPT}(\mathcal I)>0}}
    \frac{\texttt{ALG}(\mathcal I)}
         {\texttt{OPT}(\mathcal I)}.
\]
We say that $\texttt{ALG}$ is $\alpha$-competitive if $\operatorname{CR}_{\texttt{ALG}}(d)\ge\alpha(d)$ for every $d\in\mathbb N^+$. Our goal is to characterize the best achievable dependence of this ratio on $d$ and, in the extensions, on additional model parameters.

\subsection{Aggregate Marginal Information}
\label{subsec:population}

Although the horizon and future candidate profiles are unknown, the DM may have access to aggregate information about the candidate population. We formalize this information as follows.

\begin{assumption}[{\sc Exact Aggregate Marginal Advice}]
\label{assump:marginal}
Before the arrival process begins, the DM observes
\[
    \phi_k([n])
    =
    \sum_{i\in[n]}t_{ik},
    \qquad k\in[d].
\]
\end{assumption}

This advice is substantially coarser than full information about future arrivals. It consists of only $d$ aggregate totals and reveals neither which of the potentially $2^d$ binary profiles will occur nor when individual candidates will arrive. Such information may be available when a candidate pool is assembled or registered in advance, and a platform releases aggregate attribute counts, while withholding joint profiles for privacy or operational reasons. The order in which candidates become available may nevertheless remain unknown or adversarial.

In settings where the candidate pool is not known exactly, marginal totals may instead be estimated from historical records, demographic data, or forecasts. Standard concentration arguments suggest that such estimates can have small relative errors when the underlying population is large \citep{wainwright2019high}. These estimates do not, however, constitute the exact advice assumed here. We use exact marginal advice as an informational benchmark that isolates the value of coarse population information under adversarial arrivals. Related online fair-allocation models similarly use aggregate predictions \citep{banerjee2022online,banerjee2022proportionally,barman2022universal}. Section~\ref{sec:impossible} examines the consequences of misspecifying these marginal totals.

\subsection{The Value and Limits of Marginal Information}
\label{sec:impossible}

We establish two limits on what an online policy can achieve with aggregate marginal information. The first shows that exact advice is necessary to improve on the \(1/d\) dependence attainable without advice. The second shows that even exact marginal advice cannot yield a competitive ratio better than order \(1/\sqrt d\).

\paragraph{Inaccurate or absent advice.}
We first relax Assumption~\ref{assump:marginal}. Instead of observing the exact marginal count \(\phi_k([n])\), the decision maker observes a prediction \(\widehat{\phi}_k\) satisfying
\begin{align}
\label{eq:advice_error}
    \frac{\phi_k([n])}{\underline e}
    \le
    \widehat{\phi}_k
    \le
    \bar e\,\phi_k([n]),
    \qquad k\in[d],
\end{align}
where the error factors \(\underline e,\bar e\ge1\) are known. Exact advice corresponds to \(\underline e=\bar e=1\).

\begin{proposition}
\label{prop:general_impossible}
For any \(\underline e,\bar e\ge1\) with \(\underline e\bar e>1\), there exists a family of instances with inaccurate advice satisfying \eqref{eq:advice_error} for which no online policy achieves a competitive ratio better than \(O(1/d)\).
\end{proposition}

The proof, provided in Appendix~\ref{subsec:proof_general_impossible}, constructs instances with an identical arrival prefix and an identical prediction vector but different dimensions that must be protected early. Because the policy cannot distinguish these instances when making its early decisions, at least one dimension receives only an \(O(1/d)\) fraction of the offline value.

This bound is tight up to a constant. Without using any advice, a policy can draw a dimension \(J\) uniformly from \([d]\) and select the first \(K\) candidates possessing attribute \(J\). For every dimension \(k\in[d]\), this policy satisfies
\(
    \mathbb E[\phi_k(\mathcal A)]
    \ge
    \frac{1}{d}\min\{K,\phi_k([n])\}.
\)
Since
\(
    \texttt{OPT}(\mathcal I)
    \le
    \min_{k\in[d]}
    c_k\min\{K,\phi_k([n])\},
\)
the policy is \(1/d\)-competitive. It also applies when predictions are available, simply by ignoring them. Consequently, the optimal worst-case rate is \(\Theta(1/d)\) both without advice and under the multiplicative-error model in \eqref{eq:advice_error}.

Proposition~\ref{prop:general_impossible} reveals a sharp worst-case discontinuity: allowing an arbitrarily small nonzero multiplicative error can reduce the best guarantee from order \(1/\sqrt d\), obtained below with exact advice, to order \(1/d\). This differs from robustness results for Nash welfare with public or private goods \citep{banerjee2022proportionally,banerjee2022online} and for max-min fairness with private goods \citep{huang2025long}. The proposition is a worst-case theoretical result; it does not imply that every small estimation error causes such a loss in practice. Rather, it shows that a uniform guarantee beyond the \(1/d\) rate is impossible when errors and arrivals may be adversarially aligned.

\paragraph{Exact marginal advice.}
We next identify an upper bound on the performance when the marginal counts are known exactly. Although this advice reveals the total coverage of each dimension, it does not reveal how dimensions are combined within candidates. This remaining uncertainty gives rise to the following bound.

\begin{theorem}
\label{thm:impossible}
There exists a family of instances satisfying Assumption~\ref{assump:marginal} such that no online policy achieves a competitive ratio better than \(O(1/\sqrt d)\).
\end{theorem}

The proof is provided in Appendix~\ref{subsec:proof_speical_imossible}. It constructs a family of instances with the same marginal counts and a common arrival prefix, but different ways of combining dimensions within later candidates. Any policy must distribute its early selections across \(\sqrt d\) indistinguishable possibilities, which leaves some dimension with only an \(O(1/\sqrt d)\) fraction of the offline value.

Theorem~\ref{thm:impossible} shows that exact marginal advice cannot eliminate the dependence on \(d\). The policy developed in Section~\ref{sec:general} achieves an \(\Omega(1/\sqrt d)\) competitive ratio and therefore matches this impossibility bound up to a constant factor.

\subsection{Reduction to an Online Fractional Algorithm}
\label{sec:lp}

In this section, we show that because our diversity objective is evaluated using expected dimension utilities, an online fractional solution can be converted into a feasible randomized selection policy without any loss in objective value.

Suppose that, when candidate \(i\) arrives, an online fractional algorithm assigns a value \(x_i\in[0,1]\), and that the resulting vector satisfies
\(
    \sum_{i=1}^n x_i\le K.
\)
If an online rounding scheme selects candidate \(i\) with marginal probability \(x_i\), then, for every dimension \(k\in[d]\),
\[
    \mathbb E[\phi_k(\mathcal A)]
    =
    \sum_{i=1}^n \Pr[i\in\mathcal A]t_{ik}
    =
    \sum_{i=1}^n x_it_{ik}.
\]
Consequently, preserving the selection marginals preserves every expected dimension utility and hence the max-min objective.

For a fractional vector known in advance, standard dependent rounding can produce a random set of size at most \(K\) while preserving every marginal probability \citep{gandhi2006dependent}. When the fractional decisions are revealed sequentially, the online level-set rounding scheme of \citet{srinivasan2023online} provides an online implementation of the same marginal-preserving reduction.

\begin{lemma}[\sc Online Level-Set Rounding]
\label{lemma:prophet}
Suppose an online fractional algorithm assigns \(x_i\in[0,1]\) when candidate \(i\) arrives and the resulting vector \(\mathbf x\) is feasible for Problem~\eqref{eq:fluid}. Online level-set rounding makes an irrevocable selection decision for each candidate and produces a random set \(\mathcal A\) satisfying
\[
    |\mathcal A|\le K
    \quad\text{almost surely},
    \qquad
    \Pr[i\in\mathcal A]=x_i,
    \quad i\in[n].
\]
Therefore,
\[
    \emph{\texttt{ALG}}(\mathcal I)
    =
    \min_{k\in[d]}
    c_k\sum_{i=1}^n x_it_{ik}
    =
    \emph{\texttt{LU}}_{\mathcal I}(\mathbf x).
\]
\end{lemma}

Lemma~\ref{lemma:prophet} reduces the online selection problem, without loss in competitive ratio, to constructing a feasible fractional solution to Problem~\eqref{eq:fluid} online. Section~\ref{sec:general} develops such a fractional algorithm using the aggregate marginal advice.

\section{Max-Min Fairness with Marginal Information}
\label{sec:general}
Under the exact aggregate marginal advice in Assumption~\ref{assump:marginal}, the DM observes $\{\phi_k([n])\}_{k\in[d]}$ before the horizon begins, while the arrival order and the joint composition of candidate types remain adversarial. By Lemma~\ref{lemma:prophet}, it suffices to construct, online, a feasible fractional solution to Problem~\eqref{eq:fluid} with a large objective value. In this section, we first present such an online fractional algorithm and then analyze its competitive ratio.

\subsection{Online Fractional Algorithm}\label{population-fractional-algorithm}
We use a principal--agent design. Because $\texttt{OPT}(\mathcal I)$ is unknown, the principal creates one auxiliary sub-policy, which we call an \emph{agent}, for each geometric guess of $\texttt{OPT}(\mathcal I)$. Each agent runs the same two-stage policy. \emph{Controlled greedy} allocates fractional capacity to candidates that cover sufficiently many underrepresented dimensions, while \emph{minimalist protection} uses the marginal counts to protect dimensions at risk of missing the target.

A naive combination of the \(O(\log d)\) agents would lose a factor of \(\log d\). The principal avoids this loss by coupling budget and selectivity. In particular, an agent associated with a larger guess receives more budget but must also be proportionally more selective. The principal then takes the capped sum of the agents' decisions. The agents' budgets sum to \(K\), so the resulting fractional solution is feasible, while the decisions of the agent whose guess is closest to $\texttt{OPT}(\mathcal I)$ are preserved dimensionwise.

\subsubsection[Guessing OPT]{Guessing $\texttt{OPT}(\mathcal{I})$.}
\label{subsubsec:guess}
\begin{lemma}
    \label{lemma:opt_bound}
    Let
    \[
        \underline{\emph{\texttt{OPT}}}
        \triangleq \min_{k\in[d]} c_k\min\{K/d,\phi_k([n])\}
        \quad\text{and}\quad
        \overline{\emph{\texttt{OPT}}}
        \triangleq \min_{k\in[d]} c_k\min\{K,\phi_k([n])\}.
    \]
    Then $\underline{\emph{\texttt{OPT}}}\le \emph{\texttt{OPT}}(\mathcal{I})
    \le\overline{\emph{\texttt{OPT}}}\le d\,\underline{\emph{\texttt{OPT}}}$.
    If $\underline{\emph{\texttt{OPT}}}>0$, define
    \[
        R \triangleq  \max \{ 1, \lceil \log_2(\overline{\emph{\texttt{OPT}}}/\underline{\emph{\texttt{OPT}}})\rceil \},
        \qquad
        \gamma^{(r)}\triangleq 2^{r-1}\underline{\emph{\texttt{OPT}}},\quad r\in[R].
    \]
    Then $R\le \max\{1, \lceil\log_2d\rceil \}$, and some $r^*\in[R]$ satisfies
    $\emph{\texttt{OPT}}(\mathcal{I})/2\le\gamma^{(r^*)}\le\emph{\texttt{OPT}}(\mathcal{I})$.
\end{lemma}

Lemma~\ref{lemma:opt_bound} shows that $O(\log d)$ guesses computable from the observed marginal counts suffice and that one of them is within a factor of two of the unknown optimum. Its proof is a direct comparison between the capacity and marginal-count bounds and is deferred to Appendix~\ref{subsec:proof_opt_bound}.

\subsubsection{The agent's algorithm.} 
\label{subsubsec:agent}
Fix an agent $r$ with guess $\gamma^{(r)}$. The agent uses two separate stage budgets, each equal to $B_r$. The \textit{greedy budget} controls the total controlled-greedy allocation, and the \textit{protection budget} controls the total supplemental allocation. The parameters $B_r$, $\theta_r$, and $\rho_r$ are chosen by the principal below. Given the stage budgets, target $\theta_r$, and contribution threshold $\rho_r$, the agent constructs a controlled-greedy allocation $y_i^{(r)}$, dimension-wise supplemental allocations $\mathbf{z}_i^{(r)}\triangleq[z_{ik}^{(r)}]_{k\in[d]}$, and a final fractional decision $x_i^{(r)}$.

\paragraph{Determine the controlled-greedy allocation $y_i^{(r)}$.}
\label{subsubsec:determine_y}
Before setting $y_i^{(r)}$, let
\begin{align*}
    v^{(r)}_{ik} \triangleq c_k \sum_{i' < i} y^{(r)}_{i'}t_{i'k}.
\end{align*}
For $\tau\in[0,1]$, define the dimensions that candidate $i$ would still help at greedy level $\tau$:
\begin{align}
    \label{eq:definition_t_r_i}
    \mathcal{Y}^{(r)}_i(\tau)  \triangleq \{k \in [d] \mid t_{ik}=1 \text{ and } v^{(r)}_{ik}+c_k\tau<\theta_r\}.
\end{align}
The \textit{controlled greedy} rule starts from $y_i^{(r)}=0$ and increases it continuously until the first time one of the following holds:
\[
    y_i^{(r)}=1,\qquad
    \sum_{i'\le i}y_{i'}^{(r)}=B_r,\qquad
    |\mathcal{Y}_i^{(r)}(y_i^{(r)})|<\rho_r.
\]
The second condition means that the greedy budget is exhausted. Thus, controlled greedy allocates only while the current candidate covers sufficiently many dimensions that remain below the target.

\paragraph{Determine the supplemental allocations $\mathbf{z}_i^{(r)}$.}
\label{subsubsec:determine_z}
Greedy spending may still leave some dimensions exposed. The protection stage uses the marginal counts to ask whether a dimension can still reach the target using its realized utility plus all remaining candidates with that attribute. Define
\begin{align*}
w^{(r)}_{ik} \triangleq c_k \sum_{i' \le i} y^{(r)}_{i'} \cdot t_{i'k} + c_k \sum_{i' < i} z^{(r)}_{i'k} \cdot t_{i'k},
\end{align*}
and
\begin{align*}
\texttt{Res}_{ik}\triangleq c_k \sum_{i' > i} t_{i'k}
= c_k\Bigl[\phi_k([n])-\sum_{i'\le i} t_{i'k}\Bigr]
\end{align*}
for the remaining utility available to dimension $k$. If the current supplemental allocation for dimension $k$ is $\tau$, its maximal attainable utility is
\begin{align}
\label{eq:definition_m}
    m^{(r)}_{ik}(\tau) \triangleq w^{(r)}_{ik} + c_k \tau t_{ik} +\texttt{Res}_{ik}.
\end{align}
The \textit{minimalist protection} rule scans dimensions $k=1,\ldots,d$. For each $k$, it starts from $z_{ik}^{(r)}=0$ and increases $z_{ik}^{(r)}$ continuously until the first time one of the following holds:
\[
    z_{ik}^{(r)}=t_{ik},\qquad
    \sum_{i'<i}\sum_{k'\in[d]}z_{i'k'}^{(r)}+\sum_{k'\le k}z_{ik'}^{(r)}=B_r,\qquad
    m_{ik}^{(r)}(z_{ik}^{(r)})\ge\theta_r.
\]
The second stopping condition means that the protection budget is exhausted. This stage spends its protection budget only on dimensions whose current-plus-future potential is below the target, and it stops protecting a dimension as soon as that potential is restored.

Finally, agent $r$ combines $y_i^{(r)}$ and $\mathbf{z}_i^{(r)}$ by setting
\begin{align}
\label{eq:general_x_run_i}
    x_i^{(r)}
    \triangleq
    \Bigl[y_i^{(r)}+\max_{k\in[d]}\{z_{ik}^{(r)}\}\Bigr]/2.
\end{align}
Although both two-stage budgets are $B_r$, they are auxiliary and not added directly. The average in Eq.~\eqref{eq:general_x_run_i} lets the agent use a greedy budget of $B_r$ and a protection budget of $B_r$ while still spending at most $B_r$ in its final fractional vector.

\subsubsection{The principal's algorithm.}
\label{subsubsec:combine_scores}
The principal chooses the agents' parameters without splitting the capacity evenly across the $R$ guesses. At a high level, larger guesses receive more budget but must also be proportionally more selective. Specifically, for each $\gamma^{(r)}$, define
\begin{align}
\label{eq:definition_weight}
    \rho_r \triangleq \sqrt{\gamma^{(r)}/\underline{\texttt{OPT}}}=2^{(r-1)/2},
    \qquad
    W \triangleq \sum_{r=1}^{R}\rho_r .
\end{align}
Since the weights form a geometric series and $2^{R}\le 2\overline{\texttt{OPT}}/\underline{\texttt{OPT}}\le2d$,
\begin{align}
\label{eq:bound_weight}
    W = \frac{2^{R/2}-1}{\sqrt{2}-1}
    \le (2+\sqrt{2})\sqrt{d}.
\end{align}
For each agent $r$, the principal sets
\[
    \theta_r\triangleq\gamma^{(r)}/W,
    \qquad
    B_r\triangleq\rho_rK/W.
\]
Here, $\theta_r$ is the agent's target, and each of its controlled-greedy and protection stages receives a separate budget of size $B_r$. The agent uses $\rho_r$ as its greedy contribution threshold. The square-root weight $\rho_r$ balances the agent's target with the coverage supplied by its greedy budget. Indeed, since $\underline{\texttt{OPT}}\le c_kK/d$ and $\gamma^{(r)}=\rho_r^2\underline{\texttt{OPT}}$, we have
\[
    \theta_r/c_k\le\rho_r^2K/(dW).
\]
Thus, the total fractional coverage needed to bring all $d$ dimensions to the target is at most $\rho_r^2K/W$. Meanwhile, each unit spent by controlled greedy benefits at least $\rho_r$ dimensions that remain below the target. Consequently, exhausting the greedy budget supplies at least
\[
    \rho_r B_r=\rho_r^2K/W
\]
units of such coverage, matching the preceding upper bound.

At each arrival, all agents are updated independently. The principal takes their capped sum:
\begin{align}
\label{eq:combine_trial}
    \hat{x}_i \triangleq \min \Biggl\{ 1, \ \sum_{ r=1}^{R} x_i^{(r)} \Biggr\}, \quad i\in[n].
\end{align}
For every agent,
\[
    \sum_i x_i^{(r)}
    \le\frac12\left(\sum_i y_i^{(r)}+\sum_i\sum_k z_{ik}^{(r)}\right)
    \le B_r.
\]
Therefore,
\[
    \sum_i\hat{x}_i
    \le\sum_r\sum_i x_i^{(r)}
    \le\sum_rB_r
    =K.
\]
The cap ensures $0\le\hat{x}_i\le1$, so $\hat{\mathbf{x}}$ is feasible for Problem~\eqref{eq:fluid}. Moreover, $\hat{x}_i\ge x_i^{(r)}$ for every agent $r$, so the utility of the best agent is preserved dimensionwise.

Algorithm~\ref{alg:advice_augmented} outlines the full policy for the advice-augmented scenario, where the last step applies Lemma~\ref{lemma:prophet} to convert the fractional decisions into randomized selections.

\begin{algorithm}
            \caption{Policy for the Advice-Augmented Scenario} 
            \label{alg:advice_augmented}
            \begin{algorithmic}[1]
            \REQUIRE The number of dimensions $d$, their coefficients $\{ c_k \}_{k \in [d]}$, the headcount capacity $K$, and the marginal counts $\{ \phi_k([n]) \}_{k \in [d]}$
            \ENSURE The selected candidate set $\mathcal{A}$
            \STATE Compute $\underline{\texttt{OPT}}$, $\overline{\texttt{OPT}}$, $\{\gamma^{(r)}\}_{r\in[R]}$, $\{\rho_r\}_{r\in[R]}$, $W$, $\{B_r\}_{r\in[R]}$, and $\{\theta_r\}_{r\in[R]}$.
            \FOR{each candidate arrival $i$}
            \STATE Candidate $i$ and type $\mathbf{t}_i$ are revealed.
            \FOR{$r$ = 1 to $R$}
             \STATE Determine the controlled-greedy allocation $y^{(r)}_i$ with greedy budget $B_r$, target $\theta_r$, and threshold $\rho_r$.
             \STATE Determine the supplemental allocations $\mathbf{z}^{(r)}_i$ by minimalist protection with protection budget $B_r$ and target $\theta_r$.
             \STATE Set $x^{(r)}_i$ by Eq.~(\ref{eq:general_x_run_i}).
            \ENDFOR
            \STATE Take the capped sum of $\{ x^{(r)}_i \}_{r}$ as $\hat{x}_i$ using Eq.~(\ref{eq:combine_trial}).
            \STATE Apply online level-set rounding \citep{srinivasan2023online} with marginal probability $\hat{x}_i$.
            \ENDFOR
                \end{algorithmic} 
        \end{algorithm}
        
\subsection{Performance Analysis}
In this section, we show that Algorithm~\ref{alg:advice_augmented} achieves a competitive ratio of $\Omega(1/\sqrt{d})$ for the advice-augmented scenario.
The principal's construction already establishes feasibility, so the analysis focuses on an agent $r^*$ whose guess is within a factor of two of $\texttt{OPT}(\mathcal{I})$. At a high level, we consider the following two cases separately.

If the controlled-greedy allocation exhausts its greedy budget, we count the greedy mass across the underrepresented dimensions it serves. The contribution threshold lower-bounds this quantity by \(\rho_{r^*}B_{r^*}\), while the aggregate target shortfall gives the matching upper bound. Hence, equality forces every dimension to reach the target $\theta_{r^*}$.

If the greedy budget is not exhausted, the supplemental allocations cover the remaining shortfalls. When the protection budget is slack, minimalist protection keeps each dimension's current-plus-future potential above the target, yielding the target at the end of the horizon. When the protection budget is exhausted, comparison with a scaled offline optimum shows that the total residual shortfall is at most $B_{r^*}$. Since minimalist protection never oversupplies a dimension, exhausting this budget fills every shortfall.

Thus, in either case, agent $r^*$ achieves at least $\theta_{r^*}/2=\gamma^{(r^*)}/(2W)$. The capped sum preserves this value, and the bound on $W$ gives the claimed competitive ratio. Next, we present the detailed proof.

\subsubsection{Technical lemmas.}
Let $r^*$ be such that $\texttt{OPT}(\mathcal{I})/2\le\gamma^{(r^*)}\le\texttt{OPT}(\mathcal{I})$, as guaranteed by Lemma~\ref{lemma:opt_bound}. For a vector $\mathbf{u}$, write 
\[
\phi_k(\mathbf{u})\triangleq\sum_{i\in[n]}u_it_{ik}.
\]
\begin{lemma}
\label{lemma:y_capacity_depleted}
If the greedy budget is exhausted, i.e.,
$\sum_{i\in[n]}y_i^{(r^*)}=B_{r^*}$, then
\[
    \emph{\texttt{LU}}_{\mathcal I}(\mathbf y^{(r^*)})\ge\theta_{r^*}
    \quad\text{and}\quad
    \emph{\texttt{LU}}_{\mathcal I}(\mathbf x^{(r^*)})
    \ge\frac{\theta_{r^*}}{2}
    =\frac{\gamma^{(r^*)}}{2W}.
\]
\end{lemma}

\proof{Proof of Lemma~\ref{lemma:y_capacity_depleted}.}
For each dimension $k$, define
\[
    A_k\triangleq \sum_{i\in[n]}\int_0^{y_i^{(r^*)}}\mathbbm{1}[k\in\mathcal{Y}_i^{(r^*)}(\tau)]\,d\tau .
\]
The quantity $A_k$ is the amount of greedy budget spent while dimension $k$ is still below target, so $A_k\le \theta_{r^*}/c_k$. Also,
\[
    \sum_{k\in[d]}A_k
    \le \sum_{k\in[d]}\frac{\theta_{r^*}}{c_k}
    \le \sum_{k\in[d]}\frac{\rho_{r^*}^2K}{dW}
    =\rho_{r^*}B_{r^*},
\]
where the middle inequality uses $\underline{\texttt{OPT}}\le c_kK/d$. On the other hand, while the greedy process spends from its greedy budget, the current candidate covers at least $\rho_{r^*}$ underrepresented dimensions. Since the greedy budget is exhausted,
\[
    \sum_{k\in[d]}A_k
    =\sum_{i\in[n]}\int_0^{y_i^{(r^*)}}|\mathcal{Y}_i^{(r^*)}(\tau)|\,d\tau
    \ge \rho_{r^*}\sum_{i\in[n]}y_i^{(r^*)}
    =\rho_{r^*}B_{r^*}.
\]
The above two inequalities force equality throughout. In particular, $\sum_kA_k=\sum_k\theta_{r^*}/c_k$; because $A_k\le\theta_{r^*}/c_k$ holds for every $k$, equality of the sums implies $A_k=\theta_{r^*}/c_k$ for every $k$. Since $A_k\le\phi_k(\mathbf{y}^{(r^*)})$, we have $\texttt{LU}_{\mathcal{I}}(\mathbf{y}^{(r^*)})\ge\theta_{r^*}$. Finally, Eq.~(\ref{eq:general_x_run_i}) gives $\mathbf{x}^{(r^*)}\ge\mathbf{y}^{(r^*)}/2$, so $\texttt{LU}_{\mathcal{I}}(\mathbf{x}^{(r^*)})\ge\theta_{r^*}/2$.
\hfill \Halmos

The second lemma covers the complementary case, where controlled greedy leaves unused greedy budget and minimalist protection fills the remaining shortfall.
\begin{lemma}
\label{lemma:y_capacity_remain}
If the greedy budget is not exhausted, i.e.,
$\sum_{i\in[n]}y_i^{(r^*)}<B_{r^*}$, then
\[
    \emph{\texttt{LU}}_{\mathcal I}(\mathbf x^{(r^*)})
    \ge\frac{\theta_{r^*}}{2}
    =\frac{\gamma^{(r^*)}}{2W}.
\]
\end{lemma}

\proof{Proof of Lemma~\ref{lemma:y_capacity_remain}.}
Let $Z\triangleq\sum_{i\in[n]}\sum_{k\in[d]}z_{ik}^{(r^*)}$ be the total supplemental mass.

\noindent\textbf{Case 1: Slack protection budget ($Z<B_{r^*}$).}
The protection budget never binds. For notational convenience, define
\[
    P_{0k}\triangleq c_k\phi_k([n]),
    \qquad
    P_{ik}\triangleq m_{ik}^{(r^*)}(z_{ik}^{(r^*)}),
    \quad i\in[n].
\]
We claim that
\[
    P_{ik}\ge\theta_{r^*},
    \qquad i=0,1,\ldots,n,\quad k\in[d].
\]
The base case follows from
$P_{0k}=c_k\phi_k([n])\ge\texttt{OPT}(\mathcal{I})
\ge\gamma^{(r^*)}\ge\theta_{r^*}$. If $P_{i-1,k}\ge\theta_{r^*}$,
then
\[
    m_{ik}^{(r^*)}(0)
    =P_{i-1,k}+c_ky_i^{(r^*)}t_{ik}-c_kt_{ik}
    \ge \theta_{r^*}-c_kt_{ik}.
\]
Since the protection budget does not bind, minimalist protection stops either because $m_{ik}^{(r^*)}(z_{ik}^{(r^*)})\ge\theta_{r^*}$ or because the coordinate cap $z_{ik}^{(r^*)}=t_{ik}$ is reached. The latter also gives $m_{ik}^{(r^*)}(z_{ik}^{(r^*)})\ge\theta_{r^*}$. Hence the claim holds by induction. At the end of the horizon, the residual term is zero, so
\[
   m^{(r^*)}_{nk}(z^{(r^*)}_{nk}) =  c_k\left(\phi_k(\mathbf{y}^{(r^*)})+\sum_i z_{ik}^{(r^*)} \cdot t_{ik}\right)\ge\theta_{r^*}.
\]
Using Eq.~(\ref{eq:general_x_run_i}),
\[
    c_k\phi_k(\mathbf{x}^{(r^*)})
    \ge \frac{c_k}{2}\left(\phi_k(\mathbf{y}^{(r^*)})+\sum_i z_{ik}^{(r^*)} \cdot t_{ik} \right)
    \ge \theta_{r^*}/2.
\]

\noindent\textbf{Case 2: Exhausted protection budget ($Z=B_{r^*}$).}
Let
\[
    s_k\triangleq \left(\frac{\theta_{r^*}}{c_k}-\phi_k(\mathbf{y}^{(r^*)})\right)^+
\]
be the residual capacity needed by dimension $k$ after the greedy stage.

\paragraph{Step 1: Bound the aggregate shortfall.}
We first show $\sum_k s_k\le B_{r^*}$. Let $\mathbf{x}^*$ be an optimal solution to Problem~(\ref{eq:fluid}) and let $\mathcal{Q}\triangleq\{k:c_k\phi_k(\mathbf{y}^{(r^*)})<\theta_{r^*}\}$. Since $\gamma^{(r^*)}\le\texttt{OPT}(\mathcal{I})$, for every $k\in\mathcal{Q}$,
\[
    \frac{\theta_{r^*}}{c_k}
    \le \frac{\phi_k(\mathbf{x}^*)}{W}.
\]
Moreover, because the greedy budget is not exhausted, any candidate $i$ with $y_i^{(r^*)}<x_i^*/W$ could not have stopped because of the candidate cap or the greedy budget. It must therefore have stopped with $|\mathcal{Y}_i^{(r^*)}(y_i^{(r^*)})|<\rho_{r^*}$. Every dimension in $\mathcal{Q}$ remains below the target after the entire greedy stage and was therefore also below the target when the allocation to candidate $i$ stopped. Consequently, every dimension in $\mathcal{Q}$ supported by candidate $i$ belongs to $\mathcal{Y}_i^{(r^*)}(y_i^{(r^*)})$. It follows that
\begin{align*}
    & \sum_{k\in[d]}s_k
    \le \sum_{k\in\mathcal{Q}}\left(\frac{\phi_k(\mathbf{x}^*)}{W}-\phi_k(\mathbf{y}^{(r^*)})\right) = \sum_{i \in [n]} \left(\frac{x_i^*}{W}-y_i^{(r^*)}\right) \cdot \sum_{k \in \mathcal{Q}} t_{ik} \\
    &\  \le \sum_{i\in[n]}\left(\frac{x_i^*}{W}-y_i^{(r^*)}\right)^+|\mathcal{Y}_i^{(r^*)}(y_i^{(r^*)})|
    \le \rho_{r^*}\sum_{i\in[n]}\frac{x_i^*}{W}
    \le \rho_{r^*}K/W
    =B_{r^*}.
\end{align*}

\paragraph{Step 2: Show that protection never oversupplies a dimension.}
The supplemental allocation for any dimension never exceeds its residual shortfall:
\[
    \sum_i z_{ik}^{(r^*)}\le s_k,\qquad k\in[d].
\]
Indeed, if $k\notin\mathcal{Q}$, then the greedy vector alone reaches the target. Its current-plus-future potential is therefore always at least $\theta_{r^*}$, so no supplemental allocation is made for dimension $k$.

Now fix $k\in\mathcal{Q}$, and let $\nu_k$ be the first arrival such that
\[
    c_k\left(\sum_{j\le \nu_k}y_j^{(r^*)}t_{jk}
    +\sum_{j>\nu_k}t_{jk}\right)<\theta_{r^*}.
\]
Such an arrival exists because the expression equals $c_k\phi_k(\mathbf{y}^{(r^*)})<\theta_{r^*}$ at the horizon. Before $\nu_k$, the current-plus-future potential is at least the target, so $z_{ik}^{(r^*)}=0$. At arrival $\nu_k$, we thus have $m_{\nu_k,k}^{(r^*)}(0)<\theta_{r^*}$. The protection rule increases this quantity continuously and stops upon reaching $\theta_{r^*}$, or earlier if the coordinate cap or the protection budget binds. Hence,
\[
    m_{\nu_k,k}^{(r^*)}(z_{\nu_k,k}^{(r^*)})\le\theta_{r^*}.
\]
This inequality persists at every subsequent arrival. Indeed, if it holds after arrival $i-1$, then
\[
    m_{ik}^{(r^*)}(0)
    =m_{i-1,k}^{(r^*)}(z_{i-1,k}^{(r^*)})
      +c_k y_i^{(r^*)}t_{ik}-c_kt_{ik}
    \le\theta_{r^*}.
\]
Here, $-c_kt_{ik}$ removes candidate \(i\)’s contribution from the remaining-utility term, while $c_ky_i^{(r^*)}t_{ik}$ credits the fraction assigned to it by the greedy stage. Since $0\le y_i^{(r^*)}\le1$, their net contribution is nonpositive. Starting from this value, the protection rule again stops at $\theta_{r^*}$ or earlier, so $m_{ik}^{(r^*)}(z_{ik}^{(r^*)})\le\theta_{r^*}$. Induction therefore gives this bound through the final arrival. Since no residual candidates remain at the horizon,
\[
    m_{nk}^{(r^*)}(z_{nk}^{(r^*)})
    =c_k\left(\phi_k(\mathbf{y}^{(r^*)})+\sum_i z_{ik}^{(r^*)}\right)
    \le\theta_{r^*}.
\]
Rearranging gives $\sum_i z_{ik}^{(r^*)}\le s_k$, as required.

\paragraph{Step 3: Use budget exhaustion to force equality.}
Since $Z=B_{r^*}$, the inequalities
\[
    B_{r^*}=Z=\sum_k\sum_i z_{ik}^{(r^*)}\le\sum_k s_k\le B_{r^*}
\]
are all tight. Hence $\sum_i z_{ik}^{(r^*)}=s_k$ for every $k$, and so
\[
    c_k\left(\phi_k(\mathbf{y}^{(r^*)})+\sum_i z_{ik}^{(r^*)}\right)\ge\theta_{r^*}.
\]
The same comparison with Eq.~(\ref{eq:general_x_run_i}) gives $c_k\phi_k(\mathbf{x}^{(r^*)})\ge\theta_{r^*}/2$ for every $k$.
\hfill \Halmos

\subsubsection{Main result.} 
\begin{theorem}
\label{thm:population_diversity}
For every advice-augmented instance $\mathcal I$ with $\emph{\texttt{OPT}}(\mathcal I)>0$, Algorithm~\ref{alg:advice_augmented} satisfies
\[
    \frac{\emph{\texttt{ALG}}(\mathcal I)}
         {\emph{\texttt{OPT}}(\mathcal I)}
    \ge \frac{1}{4(2+\sqrt{2})\sqrt d}.
\]
Consequently, the algorithm has a competitive ratio of at least $1/[4(2+\sqrt{2})\sqrt d]$.
\end{theorem}

\proof{Proof of Theorem~\ref{thm:population_diversity}.}
The principal's construction in Section~\ref{subsubsec:combine_scores} establishes that $\hat{\mathbf{x}}$ is feasible for Problem~\eqref{eq:fluid}.
By Lemmas~\ref{lemma:y_capacity_depleted} and~\ref{lemma:y_capacity_remain},
\(
    \texttt{LU}_{\mathcal{I}}(\mathbf{x}^{(r^*)})
    \ge \frac{\gamma^{(r^*)}}{2W}
    \ge \frac{\texttt{OPT}(\mathcal{I})}{4W}.
\)
Since $\hat{\mathbf{x}}\ge\mathbf{x}^{(r^*)}$, monotonicity gives
\(
    \texttt{LU}_{\mathcal{I}}(\hat{\mathbf{x}})
    \ge \frac{\texttt{OPT}(\mathcal{I})}{4W}
    \ge \frac{\texttt{OPT}(\mathcal{I})}{4(2+\sqrt{2})\sqrt{d}},
\)
where the last inequality uses Eq.~(\ref{eq:bound_weight}). Lemma~\ref{lemma:prophet} then transfers this fractional guarantee to the randomized rounding.
\hfill \Halmos

Together with Proposition~\ref{prop:general_impossible}, Theorem~\ref{thm:population_diversity} quantifies the value of exact marginal advice. With any nonzero multiplicative error, the worst-case competitive ratio can be no better than $O(1/d)$, whereas exact advice allows Algorithm~\ref{alg:advice_augmented} to achieve $\Omega(1/\sqrt d)$. Moreover, Theorem~\ref{thm:impossible} shows that no policy can achieve a competitive ratio better than $O(1/\sqrt d)$ even with exact aggregate advice. Algorithm~\ref{alg:advice_augmented} is therefore order-optimal, matching the impossibility bound up to a constant factor.

\subsection{Extension to Nonbinary Attributes}
\label{subsec:nonbinary}

We now extend the preceding max-min analysis by allowing attributes to take nonnegative real values, with every positive value lying in a known interval \([\ell,u]\):
\[
    t_{ik}\in\{0\}\cup[\ell,u],\qquad 0<\ell\le u,
    \qquad \kappa\triangleq u/\ell,
\]
while keeping the same capacity constraint and max-min objective. The natural analog of Assumption~\ref{assump:marginal} is exact weighted marginal advice. Before the horizon, the DM knows
\(
    S_k\triangleq\sum_{i\in[n]}t_{ik}
\)
for every dimension \(k\). 

The policy in Section~\ref{population-fractional-algorithm} extends as follows. Controlled greedy replaces active-set cardinality with the weighted coverage
\(
    \sum_{k\in\mathcal Y_i^{(r)}(\tau)} t_{ik}/ \ell.
\)
In the protection stage, $z_{ik}^{(r)}\in[0,1]$ represents a supplemental selection fraction when $t_{ik}>0$, and $z_{ik}^{(r)}=0$ when $t_{ik}=0$. Minimalist protection tracks the exact residual attribute mass
\(
    S_k-\sum_{j\le i}t_{jk}
\)
and increases $z_{ik}^{(r)}$ until it reaches $1$, the protection budget is exhausted, or the target is restored. The principal's budgeting and aggregation rules are unchanged. Full details are given in Appendix~\ref{app:nonbinary}.

For any nontrivial instance with $\texttt{OPT}(\mathcal I)>0$, after normalizing the positive attributes by dividing them by $\ell$, the ratio between the upper and lower bounds computable from the weighted marginal totals on $\texttt{OPT}(\mathcal I)$ is at most $d\kappa$. Consequently, the sum of the agents' square-root weights is $O(\sqrt{d\kappa})$, which accounts for the additional $\sqrt\kappa$ factor in the guarantee below.

\begin{theorem}
\label{thm:nonbinary_positive}
For the advice-augmented scenario with nonbinary attributes, the modified policy in Appendix~\ref{app:nonbinary} has a competitive ratio of at least
\(
    1/[4(2+\sqrt{2})\sqrt{d\kappa}].
\)
\end{theorem}

\begin{theorem}
\label{thm:nonbinary_impossibility}
For the advice-augmented scenario with nonbinary attributes, there is a family of instances for which no policy can achieve a
competitive ratio better than
\(
    O\left(1 / \sqrt{d\min\{\kappa,d\}} \right).
\)
\end{theorem}

The proofs are deferred to Appendix~\ref{app:nonbinary}. When \(d=1\), the max--min objective reduces to a scalar online-allocation problem, for which the familiar single-leg revenue-management approach achieves a competitive ratio of order \(\Theta(1/(1 + \log\kappa))\) \citep{ball2009toward}. However, this logarithmic dependence in the single-dimensional case does not extend to the general multidimensional setting. For \(1\le\kappa\le d\), the two theorems above show that the optimal competitive ratio is \(\Theta(1/\sqrt{d\kappa})\). Consequently, the dependence on \(\kappa\) cannot in general be improved from square-root to logarithmic.

\section{Extension to Generalized-Mean Objectives}
\label{sec:generalized_mean}
In this section, we replace the max-min objective with generalized means while retaining binary attributes and exact aggregate marginal advice.
The generalized mean provides a natural bridge between aggregate efficiency and worst-dimension fairness. As its parameter decreases, it places increasing weight on dimensions with low utility, and as $p\to-\infty$, it recovers the max-min objective studied above. For a feasible fractional selection vector $\mathbf{x}$, let
\[
    U_k(\mathbf{x})\triangleq c_k\sum_{i\in[n]}x_it_{ik}.
\]
For $-\infty<p\le1$ and $p\ne0$, define
\[
    M_p(\mathbf{x})\triangleq
    \left(\frac1d\sum_{k=1}^d U_k(\mathbf{x})^p\right)^{1/p},
\]
where, for $p<0$, we use the convention $M_p(\mathbf{x})=0$ if $U_k(\mathbf{x})=0$ for some $k\in[d]$. For $p=0$, define
\[
    M_0(\mathbf{x})\triangleq
    \left(\prod_{k=1}^dU_k(\mathbf{x})\right)^{1/d}.
\]
For each fixed $p\in(-\infty,1]$, the \textit{generalized-mean online selection problem} retains the arrival process and capacity constraint of Section~\ref{sec:model}, but replaces the max-min objective with $M_p$.
Thus, $p=0$ gives the geometric mean, while the limit $p\to-\infty$ recovers $\min_kU_k(\mathbf{x})$. Let $\texttt{OPT}_p(\mathcal I)$ denote the maximum of $M_p$ over the feasible region of Problem~\eqref{eq:fluid}.
Lemma~\ref{lemma:prophet} continues to apply. Online level-set rounding converts a feasible fractional solution $\mathbf{x}$ into a feasible randomized selection while preserving every expected dimension utility $U_k(\mathbf{x})$, and hence its generalized mean $M_p(\mathbf{x})$. Thus, the generalized-mean online selection problem reduces to designing an online fractional algorithm that attains a large value of $M_p(\mathbf{x})$.

We first study nonnegative means $0\le p\le1$, and then turn to negative means $-\infty<p<0$.

% \subsection[Geometric Mean]{Geometric Mean: $p=0$}

\subsection{Nonnegative Means: \(0 \le p \le 1\)}

For $0\le p\le1$, the generalized mean interpolates between the geometric mean at $p=0$, also known as Nash social welfare, and the arithmetic mean at $p=1$, also known as utilitarian welfare. We first show that a logarithmic loss is inevitable over this entire range, and then present a logarithmic guarantee.

\subsubsection{Impossibility result.}
The two endpoints help reveal the source of the logarithmic barrier. When $p=1$,
\[
    M_1(\mathbf{x})
    =\frac1d\sum_{i\in[n]}x_i
      \sum_{k\in[d]}c_kt_{ik}.
\]
Thus, up to the normalization by $d$, the problem reduces to selecting candidates with scalar value $\sum_kc_kt_{ik}$. With unit coefficients, these values lie in $\{0,1,\ldots,d\}$. The marginal counts determine the total value over the horizon, but not how that value is distributed across candidates. This keeps the same informational difficulty that produces a logarithmic barrier in single-leg revenue management \citep{ball2009toward}.

At the other endpoint, $p=0$, the objective is Nash social welfare. \citet{banerjee2022proportionally} identify the same logarithmic obstruction under marginal information. More generally, their Theorem~1 establishes this obstruction throughout $0\le p\le1$. We recap their result as the following lemma.

\begin{lemma}[\sc {\citealt[Theorem~1, Adapted]{banerjee2022proportionally}}]
\label{lem:positive_generalized_mean_lower}
For every $p\in[0,1]$, there is a family of advice-augmented instances such that every online algorithm has competitive ratio at most $O(1/\log d)$ for the objective $M_p$.
\end{lemma}

\subsubsection{Our policy.}
We next give an algorithm matching the logarithmic impossibility bound. For $p=0$, \citet{banerjee2022proportionally} obtain an $\Omega(1/\log d)$ competitive ratio with a set-aside algorithm that assumes the total number of candidates is known. This information is unavailable in our setting. Their algorithm divides the capacity equally between a set-aside component that allocates an equal share to each candidate, thereby guaranteeing every dimension a baseline utility, and a static-threshold greedy component that allocates only when the marginal gain exceeds a fixed threshold.

We extend this design to all $0\le p\le1$. First, we replace the horizon-based set-aside component with one calibrated by marginal advice. Second, for $0<p\le1$, we replace the static threshold with a dynamic threshold scaled by the current $p$-mean potential. This threshold both controls the greedy budget and provides the comparison certificate used in the performance analysis. We first present and analyze the algorithm for $0<p\le1$. The endpoint $p=0$ is summarized afterward, with details deferred to Appendix~\ref{app:nash_extension}.

For $0<p\le1$, the online fractional algorithm has two components.
\paragraph{Advice-calibrated set-aside.}
For clarity, without loss of generality, we may assume that $\phi_k([n])>0$ for every dimension.
 Define
\[
    a_k\triangleq \min\left\{1,\frac{K}{d\phi_k([n])}\right\},
    \qquad
    \delta_k\triangleq \frac{c_k}{2}\phi_k([n])a_k
    =
    \frac{c_k}{2}\min\left\{\phi_k([n]),\frac Kd\right\}.
\]
When candidate $i$ arrives, the set-aside component assigns
\[
    y_i=\frac12\max_{k:t_{ik}=1}a_k .
\]
We use the convention that the maximum over an empty set is zero.

\paragraph{Dynamic-threshold greedy allocation.}
Let $z_i$ denote the additional allocation to candidate $i$. For a tentative allocation $z$ to the current candidate, define the promised utility of dimension $k$, the associated $p$-mean potential, and the marginal score by
\[
    P_{ik}(z)\triangleq
    \delta_k+c_k\sum_{j<i}z_j t_{jk}+c_kzt_{ik},
    \qquad
    \Phi_i(z)\triangleq\frac1d\sum_{k=1}^dP_{ik}(z)^p,
\]
and
\[
    \sigma_i(z)\triangleq
    \frac1d\sum_{k=1}^dc_kt_{ik}P_{ik}(z)^{p-1}.
\]
With $\eta_d\triangleq4\log(2d+2)$, increase $z_i$ continuously from zero until the first of the following events occurs: $z_i=1-y_i$, the greedy budget $K/2$ is reached, or
\[
    \sigma_i(z_i)\le\frac{\eta_d}{K}\Phi_i(z_i).
\]
The algorithm then sets $x_i=y_i+z_i$.

\begin{theorem}
\label{thm:positive_generalized_mean}
For every $p\in(0,1]$, the above algorithm is feasible and satisfies
\(
    M_p(\mathbf{x})
    \ge
    \frac{\emph{\texttt{OPT}}_p(\mathcal I)}{1+4\log(2d+2)} .
\)
After online level-set rounding, the same guarantee holds for the expected dimension utilities of the randomly selected set.
\end{theorem}

\proof{Proof of Theorem~\ref{thm:positive_generalized_mean}.}
The proof has three steps. We establish the set-aside guarantees, show that the greedy budget never binds, and compare the resulting solution with an arbitrary feasible offline solution.

\paragraph{Step 1: Establish the set-aside guarantees.}
The set-aside component uses at most half of the capacity because
\[
    \sum_i y_i
    \le
    \frac12\sum_i\sum_{k=1}^{d} t_{ik}a_k
    =
    \frac12\sum_{k=1}^{d}\phi_k([n])a_k
    \le
    \frac K2 .
\]
It also gives every dimension its promised baseline utility:
\[
    Y_k\triangleq c_k\sum_i y_it_{ik}\ge \delta_k .
\]

\paragraph{Step 2: Show that the greedy budget never binds.}
Let
\[
    P_k\triangleq \delta_k+c_k\sum_i z_it_{ik},
    \qquad
    \Phi\triangleq\frac1d\sum_{k=1}^{d}P_k^p,
    \qquad
    \Phi_0\triangleq\frac1d\sum_{k=1}^{d}\delta_k^p .
\]
During an infinitesimal greedy increase on candidate $i$, we have $\mathrm{d} \Phi/ \mathrm{d}z=p\sigma_i(z)$. Thus, while the stopping rule has not been triggered,
\[
    \frac{\mathrm{d}}{\mathrm{d}z}\log\Phi
    =
    \frac{p\sigma_i(z)}{\Phi}
    \ge
    \frac{p\eta_d}{K}.
\]
If the greedy component used all of $K/2$, integration would give
\[
    \log \frac{\Phi}{\Phi_0}\ge \frac{p\eta_d}{2}.
\]
But before the greedy budget is exhausted, each dimension satisfies a common bound $P_k\le(2d+2)\delta_k$. Indeed, if $\phi_k([n])\le K/d$, then $P_k\le\delta_k+c_k\phi_k([n])\le3\delta_k$; if $\phi_k([n])>K/d$, then $P_k\le\delta_k+c_kK/2=(d+1)\delta_k$. Hence
\[
    \log\frac{\Phi}{\Phi_0}\le p\log(2d+2),
\]
contradicting $\eta_d=4\log(2d+2)$. Therefore $\sum_i z_i<K/2$, and $\sum_i x_i=\sum_i y_i+\sum_i z_i\le K$.

\paragraph{Step 3: Compare with an offline benchmark.}
Let $\mathbf{w}$ be any feasible offline vector. Since $Y_k\ge\delta_k$, the final utilities satisfy
\[
    U_k(\mathbf{x})\ge P_k,\qquad k\in[d].
\]
For each candidate $i$, define the final marginal score
\[
    s_i\triangleq \frac1d\sum_{k=1}^{d}c_kt_{ik}U_k(\mathbf{x})^{p-1}.
\]
If $x_i<1$, then $z_i<1-y_i$. Because the greedy budget is not exhausted, candidate $i$ must have stopped by the threshold rule. Hence,
\begin{align*}
    s_i
    =
    \frac1d\sum_{k=1}^{d}c_kt_{ik}U_k(\mathbf{x})^{p-1}
    \le
    \frac1d\sum_{k=1}^{d}c_kt_{ik}P_{ik}(z_i)^{p-1}
    =
    \sigma_i(z_i)
    \le
    \frac{\eta_d}{K}\Phi_i(z_i)
    \le
    \frac{\eta_d}{K}\Phi
    \le
    \frac{\eta_d}{K}M_p(\mathbf{x})^p .
\end{align*}
Here, the first inequality follows from $U_k(\mathbf{x})\ge P_k\ge P_{ik}(z_i)$ and $p-1\le0$. The last two inequalities follow from $P_k\ge P_{ik}(z_i)$ and $U_k(\mathbf{x})\ge P_k$, respectively, because $p>0$.

Writing $r_i=(w_i-x_i)^+$, the threshold bound applies whenever $r_i>0$ because then $x_i<1$. Using $\sum_ir_i\le\sum_iw_i\le K$, we obtain the certificate
\begin{align*}
    & \frac1d\sum_{k=1}^{d}U_k(\mathbf{w})U_k(\mathbf{x})^{p-1}
    =
    \sum_iw_is_i
    \le
    \sum_ix_is_i
    +
    \frac{\eta_d}{K}M_p(\mathbf{x})^p\sum_ir_i \\
    &\ =
    M_p(\mathbf{x})^p
    +
    \frac{\eta_d}{K}M_p(\mathbf{x})^p\sum_ir_i
    \le
    (1+\eta_d)M_p(\mathbf{x})^p .
\end{align*}
The second equality uses
\[
    \sum_i x_i s_i
    =
    \frac1d\sum_{k=1}^{d}
    U_k(\mathbf{x})^{p-1}\sum_i c_kt_{ik}x_i
    =
    \frac1d\sum_{k=1}^{d}U_k(\mathbf{x})^p
    =
    M_p(\mathbf{x})^p.
\]

It remains to translate this certificate into a bound on the objective.
Every dimension has $U_k(\mathbf{x})>0$ by the set-aside guarantee.
Define
\[
    \lambda_k
    \triangleq
    \frac{U_k(\mathbf{x})^p}{dM_p(\mathbf{x})^p},
    \qquad k\in[d],
\]
so that $\sum_k\lambda_k=1$.
Since $u\mapsto u^p$ is concave for $0<p\le1$, Jensen's inequality gives
\begin{align*}
    \frac{M_p(\mathbf{w})^p}{M_p(\mathbf{x})^p}
    =
    \sum_{k=1}^{d}
    \lambda_k
    \left(\frac{U_k(\mathbf{w})}{U_k(\mathbf{x})}\right)^p
    \le
    \left(
    \sum_{k=1}^{d}
    \lambda_k
    \frac{U_k(\mathbf{w})}{U_k(\mathbf{x})}
    \right)^p.
\end{align*}
Taking the $p$th root and substituting the definition of $\lambda_k$ yields
\[
    M_p(\mathbf{w})
    \le
    M_p(\mathbf{x})^{1-p}
    \left[
    \frac1d\sum_{k=1}^{d}
    U_k(\mathbf{w})U_k(\mathbf{x})^{p-1}
    \right]
    \le
    (1+\eta_d)M_p(\mathbf{x}).
\]
Taking $\mathbf{w}$ to be an optimal solution gives the claimed fractional guarantee. Lemma~\ref{lemma:prophet} then transfers it to randomized selections because level-set rounding preserves all marginal probabilities.
\hfill \Halmos

\paragraph{The geometric mean.}
When $p=0$, the objective becomes Nash social welfare. If any dimension has zero marginal count, then $\texttt{OPT}_0(\mathcal I)=0$, so the guarantee is immediate. We therefore consider instances with positive marginal counts in every dimension. Taking the limit $p\to0$ replaces the marginal score above with
\(
    \frac1d\sum_{k=1}^d\frac{c_kt_{ik}}{P_{ik}(z)}.
\)
Using this value in the same dynamic-threshold framework gives the Nashian policy analyzed in Appendix~\ref{app:nash_extension} and the following result.

\begin{theorem}
\label{thm:geometric_mean}
For the advice-augmented scenario, the Nashian policy defined in Appendix~\ref{app:nash_extension} is feasible and satisfies
\(
    M_0(\mathbf{x})
    \ge
    \frac{\emph{\texttt{OPT}}_0(\mathcal I)}{4\log(2d+2)} .
\)
After online level-set rounding, the same guarantee holds for the expected dimension utilities of the randomly selected set.
\end{theorem}

% The proof is deferred to Appendix~\ref{app:nash_extension}. This endpoint is useful below not only as a welfare guarantee but also as a certificate: a Nashian run on any dimension set $D$ produces utilities $U_k^N$ such that every feasible comparator $\mathbf{w}$ satisfies
% \[
%     \frac1{|D|}\sum_{k\in D}\frac{U_k(\mathbf{w})}{U_k^N}
%     \le
%     \mathrm{polylog}(d),
% \]
% after the budget reduction used in the negative-mean policy.

\subsection[Negative Means]{Negative Means: $-\infty<p<0$}

Negative means place increasing weight on low-utility dimensions and are therefore qualitatively closer to the max-min objective. Throughout this subsection, write $p=-q$ with $q>0$.

\subsubsection{Impossibility result.} We first extend the proof of Theorem~\ref{thm:impossible} to obtain the following impossibility result.

\begin{theorem}
\label{thm:negative_generalized_mean_lower}
Fix $q>0$ and write $p=-q$. There is a family of advice-augmented instances such that every policy has a competitive ratio of at most
\(
    O_q\left(d^{-\frac{q}{2q+1}}\right)
\)
for $M_{-q}$.
\end{theorem}

The proof is deferred to Appendix~\ref{app:pmean_hardness}. Thus, for every fixed $p<0$, a polynomial decay in $d$ is unavoidable. This rate differs from the corresponding impossibility bounds for online private-goods allocation with predictions~\citep{huang2025long}. Writing again $p=-q$, those bounds are $O(d^{-1/2+\epsilon})$ for $q\ge1$ and $O_q(d^{-q/(q+1)+\epsilon})$ for $0<q<1$, for any $\epsilon>0$. Hence, although the two models have the same dependence on $d$ at the endpoints $p=0$ and $p=-\infty$, their rates differ for other negative means.

\subsubsection{Our policy.}
\label{subsubsec:negative_mean_policy}
We next outline an algorithm that matches Theorem~\ref{thm:negative_generalized_mean_lower} up to polylogarithmic factors. Implementation details and the complete proof are deferred to Appendix~\ref{app:nash_to_harmonic}. Its high-level architecture is the principal-agent design from Section~\ref{population-fractional-algorithm}. Recall that the principal there creates agents for geometric guesses of the unknown optimum, assigns each agent a budget, lets the agents operate independently, and combines their decisions by a capped sum. This capped sum preserves every dimension's utility achieved by the agent with the successful guess. We use the same architecture here with one additional layer.

The construction has three levels: a global guess determines capacity groups, each group is paired with a reference set and a group-specific guess, and the resulting agent combines Nashian allocation with minimalist protection.

\begin{samepage}
For every nonempty set of dimensions $D\subseteq[d]$, define the restricted negative mean and its offline benchmark by
\begin{align*}
    M_{-q}^{D}(\mathbf{x})
    &\triangleq
    \left(
    \frac1{|D|}\sum_{k\in D}U_k(\mathbf{x})^{-q}
    \right)^{-1/q},\\
    \texttt{OPT}_{-q}^{D}(\mathcal I)
    &\triangleq
    \max\left\{
    M_{-q}^{D}(\mathbf{x}):
    \sum_{i\in[n]}x_i\le K,\
    0\le x_i\le1\ \text{for all }i\in[n]
    \right\}.
\end{align*}
We use the convention $M_{-q}^{D}(\mathbf{x})=0$ if $U_k(\mathbf{x})=0$ for some $k\in D$. Thus, restricting the benchmark to $D$ only changes the dimensions included in the objective. In particular,
\(
    M_{-q}^{[d]}(\mathbf{x})=M_{-q}(\mathbf{x})
\)
and
\(
    \texttt{OPT}_{-q}^{[d]}(\mathcal I)=\texttt{OPT}_{-q}(\mathcal I).
\)
Write
\[
    T_D\triangleq\texttt{OPT}_{-q}^{D}(\mathcal I).
\]
\end{samepage}

Let
\[
    T\triangleq T_{[d]}=\texttt{OPT}_{-q}(\mathcal I),
    \qquad
    V_k\triangleq c_k\min\{\phi_k([n]),K\}.
\]
If $T=0$, the claimed competitive guarantee is immediate, so we consider the nontrivial case $T>0$. The quantity \(V_k\) is computable from the marginal count \(\phi_k([n])\) and upper-bounds the utility that any feasible solution can give dimension $k$. Let $\mathbf{x}^*$ attain $T$ and set $U_k^*\triangleq U_k(\mathbf{x}^*)$. Then
\[
    \frac1d\sum_{k=1}^d\left(\frac{T}{U_k^*}\right)^q=1,
    \qquad
    \frac1d\sum_{k=1}^d\left(\frac{T}{V_k}\right)^q\le1.
\]
These two relations are the backbone of the construction. They show that dimensions with very small \(V_k\) cannot occupy too much mass in the negative-mean objective.

\paragraph{The principal's design.}
The principal first forms a geometric grid of candidate values $\Gamma$ for the unknown global benchmark $T$. Fix the successful global guess satisfying $T/2\le\Gamma\le T$. Let $\lambda\triangleq2^{-1/q}$ and $J\triangleq\lceil\log_2d\rceil+1$. This guess partitions dimensions with comparable values of \(V_k\) into
\[
    G_{\mathrm{hi}}\triangleq\{k:V_k\ge\lambda\Gamma\},
    \qquad
    G_\ell\triangleq\{k:\lambda^{\ell+1}\Gamma\le V_k<\lambda^\ell\Gamma\},
    \quad \ell=1,\ldots,J-1 .
\]
The bound $\frac1d\sum_k(T/V_k)^q\le1$ ensures that these groups cover all dimensions. 
% Empty groups are omitted.

The principal next pairs each capacity group $G$ with a reference set $D$. These two sets serve different purposes. The set $G$ is the \emph{protected set}, containing the dimensions to which minimalist protection is applied, whereas $D$ is the \emph{reference set}, on which the Nashian stage and the restricted-objective benchmark $T_D$ are defined. They coincide for the lower-capacity groups but differ for the high-capacity group.

For the high-capacity group, the protected set is $G_{\mathrm{hi}}$, while the reference set is $D_{\mathrm{hi}}=[d]$ and its optimum is $T_{D_{\mathrm{hi}}}=T$. Using this global reference keeps the benchmark comparable to \(V_k\) of every protected dimension. For a lower-capacity group, we set $D_\ell=G_\ell$ and define
\[
    T_\ell\triangleq T_{D_\ell}
    =\texttt{OPT}_{-q}^{D_\ell}(\mathcal I)
\]
as the restricted-objective benchmark for that group. We use $T_{D_\ell}$ and the shorter notation $T_\ell$ interchangeably. These group benchmarks are also unknown, so the principal uses a second geometric grid to obtain a constant-factor guess of each $T_\ell$. For uniform notation, let $\Gamma_D$ denote the reference-optimum guess assigned to an agent, with $\Gamma_{D_{\mathrm{hi}}}=\Gamma$ and $T_\ell/2\le\Gamma_{D_\ell}\le T_\ell$ for a successful lower-group guess. Thus, each agent is specified by a global guess $\Gamma$, a capacity group induced by that guess, and a reference-optimum guess $\Gamma_D$. Appendix~\ref{app:nh_policy} shows how the principal budgets all such agents in parallel while losing only polylogarithmic factors.

\paragraph{The agent's algorithm.}
Each agent first runs the Nashian algorithm of Appendix~\ref{app:nash_extension} on its reference set $D$ with a reduced budget. Let $U_k^N$ denote the resulting utilities. It then applies minimalist protection only to dimensions in its protected set $G$ whose Nashian utility is below the floor $\beta\Gamma_D$, where the parameter $\beta>0$ is calibrated in the proof below, raising them toward this floor. This is the same minimalist process used in the max-min algorithm. It allocates only when the current utility together with all utility still available from future arrivals would otherwise be insufficient to reach the floor.

The agent's two stages have complementary roles. The Nashian certificate ensures that only a small number of dimensions fall below the floor. Meanwhile, grouping dimensions by $V_k$ guarantees $T_D=O_q(V_k)$ within each group, so protecting any one deficient dimension consumes only $O_q(\beta K)$ capacity. Their product bounds the agent's minimalist-protection budget. Finally, the principal combines all agents' decisions using the same capped sum as before.

\begin{theorem}[\sc Negative Generalized-Mean Guarantee]
\label{thm:negative_generalized_mean}
Fix $q>0$ and write $p=-q$. The negative-mean policy detailed in Appendix~\ref{app:nash_to_harmonic} and outlined above produces a feasible fractional vector $\mathbf{x}$ satisfying
\[
    M_{-q}(\mathbf{x})
    \ge
    \frac{\Omega_q(1)\,\emph{\texttt{OPT}}_{-q}(\mathcal I)}
    {d^{\,q/(2q+1)}\mathrm{polylog}(d)} .
\]
After online level-set rounding, the generalized mean of the expected dimension utilities satisfies the same guarantee.
\end{theorem}

\paragraph{Sketch of proof.}
% We suppress constants depending only on $q$ and the additional polylogarithmic loss from running the guesses in parallel. Because $T_D/2\le\Gamma_D\le T_D$ for the successful agent, we write the protection floor as $\beta T_D$ throughout the sketch. This changes only constant factors. 
The construction runs $P=O(\log^3d)$ agents, each with Nashian and protection budgets both equal to $B\triangleq K/(2P)$. We consider a successful global guess $T/2\le\Gamma\le T$ and, for each associated reference set $D$, a successful agent whose group-specific guess satisfies $T_D/2\le\Gamma_D\le T_D$.
The proof has three steps.

\paragraph{Step 1: Identify the dimensions that need protection.}
Fix a protected set $G$ and its associated reference set $D$. By adapting the arguments in Appendix~\ref{app:nash_extension} for the Nashian policy, we obtain a Nashian certificate over $D$: for every feasible allocation $\mathbf{w}$,
\[
    \frac1{|D|}\sum_{k\in D}\frac{U_k(\mathbf{w})}{U_k^N}
    \le L,
    \qquad L=\mathrm{polylog}(d).
\]
This comparison property, rather than the geometric-mean guarantee itself, is what we need. Let $\mathbf{x}^{*,D}$ attain $\texttt{OPT}_{-q}^{D}(\mathcal I)$ and set $U_k^{*,D}\triangleq U_k(\mathbf{x}^{*,D})$. Then, by the definition of $M_{-q}^{D}$ and $T_D$,
\[
    \frac1{|D|}\sum_{k\in D}
    \left(\frac{T_D}{U_k^{*,D}}\right)^q=1.
\]
Set $\xi\triangleq q/(q+1)$. Applying H\"older's inequality to the two preceding formulas gives
\begin{align*}
    \frac1{|D|}\sum_{k\in D}\left(\frac{T_D}{U_k^N}\right)^\xi
    =
    \frac1{|D|}\sum_{k\in D}
    \left(\frac{U_k^{*,D}}{U_k^N}\right)^\xi
    \left(\frac{T_D}{U_k^{*,D}}\right)^\xi
    \le
    \left(
    \frac1{|D|}\sum_{k\in D}
    \frac{U_k^{*,D}}{U_k^N}
    \right)^\xi
    \left(
    \frac1{|D|}\sum_{k\in D}
    \left(\frac{T_D}{U_k^{*,D}}\right)^q
    \right)^{1/(q+1)}
    \le L^\xi.
\end{align*}
If $U_k^N<\beta T_D$, then $(T_D/U_k^N)^\xi>\beta^{-\xi}$. Comparing this lower bound with the preceding moment bound shows that at most
\(
    |D|(\beta L)^\xi
\)
dimensions in the group require protection. Thus, the Nashian stage reduces protection to a small exceptional set.

\paragraph{Step 2: Bound the minimalist-protection budget.}
The capacity groups make each exceptional dimension inexpensive to protect. For the high-capacity group, $T\le2V_k/\lambda$ because $\Gamma\ge T/2$. For a lower-capacity group $G_\ell$,
\[
    T_\ell
    \le \max_{k' \in G_\ell}V_{k'}
    <\lambda^\ell\Gamma
    \le\frac{V_k}{\lambda},
    \qquad k\in G_\ell.
\]
Thus, in either case, $T_D\le 2V_k/\lambda$ for every protected dimension. Raising dimension $k$ to $\beta \Gamma_D$ requires at most
\[
    \frac{\beta \Gamma_D}{c_k}
    \le
    \frac{\beta T_D}{c_k}
    \le
    \frac{2\beta V_k}{\lambda c_k}
    \le
    \frac{2\beta K}{\lambda}
\]
capacity, where the last inequality uses $V_k/c_k=\min\{\phi_k([n]),K\}\le K$.

Multiplying this per-dimension cost by the count from Step~1 shows that the protection cost associated with reference set $D$ is at most
\[
    \frac{2|D|}{\lambda}\beta^{1+\xi}L^\xi K.
\]
Choose
\[
    \beta
    \triangleq
    a_q\left(\frac{\lambda}{PdL^\xi}\right)^{1/(1+\xi)},
\]
where $a_q>0$ is sufficiently small and depends only on $q$. 
Since $|D|\le d$, the preceding protection-cost bound is at most $2a_q^{1+\xi}K/P\le K/(2P)=B$. Thus, every successful agent raises all deficient dimensions to their floors without exceeding its protection budget. 
% Every agent uses at most $B$ in each of its two stages, so all $P$ agents together use at most $2PB=K$. Hence the combined fractional solution is feasible.

\paragraph{Step 3: Combine the group-level guarantees.}
After protection, the final utility of every dimension in the protected set $G$ is, up to a constant, at least both $U_k^N$ and $\beta T_D$. Since
\[
    \max\{a,b\}\ge a^{1-\xi}b^\xi
    \quad\text{and}\quad
    q(1-\xi)=\xi,
\]
we have
\[
    U_k(\mathbf{x})
    \ge
    \Omega_q\left((U_k^N)^{1-\xi}(\beta T_D)^\xi\right),
\]
and therefore
\[
    \left(\frac{T_D}{U_k(\mathbf{x})}\right)^q
    \le
    O_q\left(
    \beta^{-q\xi}
    \left(\frac{T_D}{U_k^N}\right)^\xi
    \right).
\]
Summing this inequality over $G\subseteq D$ and applying the moment bound over $D$ from Step~1 gives the local guarantee
\[
    \frac1{|D|}\sum_{k\in G}
    \left(\frac{T_D}{U_k(\mathbf{x})}\right)^q
    \le
    \Psi,
    \qquad
    \Psi\triangleq O_q\left(L^\xi\beta^{-q\xi}\right).
\]
For the high-capacity group, $D=[d]$ and $T_D=T$, so its contribution to the global inverse-power sum is at most $\Psi$. For a lower-capacity group $G_\ell$, where $D=G_\ell$ and $T_D=T_\ell$,
\[
    \frac1d\sum_{k\in G_\ell}
    \left(\frac{T}{U_k(\mathbf{x})}\right)^q
    \le
    \Psi\,
    \frac{|G_\ell|}{d}
    \left(\frac{T}{T_\ell}\right)^q.
\]
To sum these bounds, observe that the global optimizer $\mathbf{x}^*$ belongs to the same fractional feasible region used by $\texttt{OPT}_{-q}^{G_\ell}(\mathcal I)$. Therefore,
\[
    T_\ell
    =\texttt{OPT}_{-q}^{G_\ell}(\mathcal I)
    \ge M_{-q}^{G_\ell}(\mathbf{x}^*)
    =
    \left(
    \frac1{|G_\ell|}
    \sum_{k\in G_\ell}(U_k^*)^{-q}
    \right)^{-1/q},
\]
which is equivalent to
\[
    \frac{|G_\ell|}{d}\left(\frac{T}{T_\ell}\right)^q
    \le
    \frac1d\sum_{k\in G_\ell}
    \left(\frac{T}{U_k^*}\right)^q.
\]
Because the lower-capacity groups are disjoint and
$\frac1d\sum_k(T/U_k^*)^q=1$, their weights sum to at most one. Adding the high-capacity group therefore gives
\[
    \frac1d\sum_{k=1}^d\left(\frac{T}{U_k(\mathbf{x})}\right)^q
    \le 2\Psi
    \le
    d^{\,q^2/(2q+1)}\mathrm{polylog}(d),
\]
where the last inequality follows from
\[
    \beta^{-q\xi}
    =
    a_q^{-q\xi}\left(\frac{PdL^\xi}{\lambda}\right)^{q\xi/(1+\xi)}
    =
    d^{\,q^2/(2q+1)}\mathrm{polylog}(d).
\]
Finally, by the definition of $M_{-q}$,
\[
    \frac{M_{-q}(\mathbf{x})}{T}
    =
    \left[
    \frac1d\sum_{k=1}^d
    \left(\frac{T}{U_k(\mathbf{x})}\right)^q
    \right]^{-1/q},
\]
which yields the claimed rate and completes the proof sketch.
\hfill \Halmos

\section{Conclusion}
\label{sec:conclusion}
We study diversity-fair online selection under adversarial arrivals. The decision maker selects candidates subject to a shared headcount capacity, and each selected candidate may contribute to several diversity dimensions. Before arrivals begin, the decision maker observes the aggregate marginal count for each dimension, but not how dimensions overlap within candidates or when those candidates will arrive. We characterize the competitive guarantees attainable with this information when dimension utilities are evaluated by generalized means over the full range $-\infty\le p\le1$.

For max-min fairness, our controlled-greedy and minimalist-protection algorithm achieves an $\Omega(1/\sqrt d)$ competitive ratio, and a matching impossibility result establishes the optimal dependence on $d$ up to a constant factor. The principal-agent design combines geometric guesses of the offline optimum using nonuniform budgets, thereby avoiding an additional logarithmic loss. Exact aggregate marginal advice improves the optimal rate from $\Theta(1/d)$ without advice or under misspecified advice to $\Theta(1/\sqrt d)$. For nonbinary attributes, we also obtain matching bounds when $1\le\kappa\le d$.

The generalized-mean results reveal how the attainable guarantee changes with the fairness objective. For $0\le p\le1$, the optimal competitive ratio is $\Theta(1/\log d)$. For each fixed finite negative mean $p=-q<0$, our algorithm achieves $d^{-q/(2q+1)}$ up to polylogarithmic factors, matching the polynomial dependence in the impossibility bound. As $q$ increases, the exponent $q/(2q+1)$ approaches the tight exponent $1/2$ for max-min fairness. Thus, greater emphasis on the least-served dimensions produces a gradual transition from logarithmic to polynomial dependence on $d$.

We conclude with three directions for future research. First, for each fixed $q>0$, can the polylogarithmic loss for $p=-q$ be removed to obtain the tight rate $\Theta_q(d^{-q/(2q+1)})$? This requires determining whether the losses from the Nashian certificate, geometric guesses, and dimension grouping are artifacts of the current analysis or are unavoidable. Second, what alternative forms of advice can support online selection? In divisible online allocation, \citet{cohen2023general} obtain nearly optimal guarantees using one learned proportional-allocation parameter per agent. It remains to determine whether analogous parameters for diversity dimensions can replace aggregate marginal counts, whether they can be learned efficiently, and how estimation error affects the resulting guarantees. The overlapping benefits of candidate selection make such an extension nontrivial.
% Finally, under what structured forms of uncertainty can advice continue to
    % improve worst-case guarantees? Promising models include bounded additive
    % errors relative to capacity, interval-valued advice, partial informationutilities across several dimensions
    % about joint candidate profiles, and stochastic or random-order arrivals.,
    % A central goal is to characterize a gradual relationship between advice
    % quality and the competitive ratio, rather than the collapse caused by
    % unrestricted misspecification.

% \section*{Acknowledgment}
% Authors are ordered alphabetically.
 % and Tongwen Wu is the corresponding author.

    \bibliographystyle{informs2014}
\bibliography{sample}

 \ECSwitch
 \def\theHsection{EC.\arabic{section}}
\def\theHsubsection{\theHsection.\arabic{subsection}}
\def\theHsubsubsection{\theHsubsection.\arabic{subsubsection}}
\def\theHequation{EC.\arabic{equation}}
\def\theHlemma{EC.\arabic{lemma}}

\ECHead{\centering Electronic Companions  \\to ``Diversity-Fair Online Selection''}

\section{Omitted Details in Section~\ref{sec:model}}
\subsection{Discussion on Offline Intractability}
\label{subsec:intract}
As mentioned in Section~\ref{subsubsec:diversity}, we provide a detailed discussion on the intractability of two alternative modeling approaches. In this subsection, we focus exclusively on the offline problem. To show the intractability, we rely on the reduction from an NP-hard problem presented below.

\textit{3-dimensional matching (3DM).} An instance of 3DM consists of a hypergraph $G = (\mathcal{X}, \mathcal{Y}, \mathcal{Z}, \mathcal{E})$, where $\mathcal{X}, \mathcal{Y}$, and $\mathcal{Z}$ are disjoint sets of vertices of size $m$, and the edge set $\mathcal{E}$ is a subset of $\mathcal{X} \times \mathcal{Y} \times \mathcal{Z}$. A matching $\mathcal{M}$ is a subset of $\mathcal{E}$ such that any two edges in this subset are disjoint. The 3DM problem asks whether there exists a perfect matching that covers all the vertices (i.e., $|\mathcal{M}| = m$), which is NP-hard (Johnson and Garey 1979).
% \hfill $\clubsuit$

First, for the deterministic problem, $\max_{\mathcal{A} \subseteq [n]:|\mathcal{A}| \le K} \min_{k \in [d]} c_k \phi_k(\mathcal{A})$, we show that finding an $\alpha$-approximate solution for any constant factor $\alpha > 0$ is NP-hard. A solution $\mathcal{A}$ is considered $\alpha$-approximate if $\min_{k \in [d]} c_k \phi_k(\mathcal{A}) \ge \alpha \min_{k \in [d]} c_k \phi_k(\mathcal{A}^*)$, where $\mathcal{A}^*$ is the optimal solution. To establish this, we show any instance of the 3DM problem can be reformulated as a selection problem with $d = 3m$ dimensions corresponding to the vertices  $\mathcal{X} \cup \mathcal{Y} \cup \mathcal{Z}$, all coefficients $c_k$ set as 1, and $|\mathcal{E}|$ candidates where each one corresponds to one edge and has 3 attributes corresponding to the vertices of the edge. Finding a perfect matching in 3DM is equivalent to selecting a set of candidates, $\mathcal{A}$, subject to the capacity constraint $|\mathcal{A}| \le m$, such that $\min_{k \in [d]} \phi_k(\mathcal{A}) \ge 1$. Therefore, for any constant $\alpha > 0$, it is NP-hard to find an $\alpha$-approximate solution. Otherwise, for the instance with a perfect matching, we can find a solution $\hat{\mathcal{A}}$ with $\min_{k \in [d]} \phi_k(\hat{\mathcal{A}}) \ge \alpha$, implying that $\hat{\mathcal{A}}$ would be a perfect matching (since the objective values are integers). Hence, the deterministic problem is computationally intractable and unsuitable for online settings.

Second, we examine the problem using a randomized algorithm to maximize $\mathbb{E}[\min_{k \in [d]} c_k \phi_k(\mathcal{A})]$. Here, an $\alpha$-approximation algorithm must satisfy $\mathbb{E}[\min_{k \in [d]} c_k \phi_k(\mathcal{A})] \ge \alpha \min_{k \in [d]} c_k \phi_k(\mathcal{A}^*)$, for any constant $\alpha > 0$. We argue that deriving such an algorithm is highly challenging. For any instance of 3DM and its corresponding selection problem, since each solution $\mathcal{A}$ can only have an objective value of either zero or one (because the size of $\mathcal{A}$ is at most $m$, and each candidate contributes to exactly three dimensions out of $3m$), the optimal solution $\mathcal{A}^*$ achieves $\min_{k \in [d]} c_k \phi_k(\mathcal{A}^*) = 1$ if a perfect matching exists. Therefore, if an $\alpha$-approximation algorithm existed, it could find a perfect matching with probability $\alpha$ if one exists. Given the NP-hardness of 3DM, it is unlikely that such an $\alpha$-approximation algorithm exists for the randomized version of the problem.

\subsection{Proof Lemma~\ref{lemma:equivalence}.}
\label{subsec:proof_equiv}
% Throughout the proof of this proposition, we consider the algorithm as the optimal offline algorithm. 
To prove that $\texttt{OPT}(\mathcal{I})$ is equal to the performance $\texttt{OFF}(\mathcal{I})$ of the optimal offline algorithm, we first establish that $\texttt{OPT}(\mathcal{I}) \ge \texttt{OFF}(\mathcal{I})$, and then demonstrate that $\texttt{OPT}(\mathcal{I}) \le \texttt{OFF}(\mathcal{I})$.

We begin with the first inequality. Let $x'_i$ represent the ex-ante selection probability of candidate $i \in [n]$ in the optimal offline algorithm. Since the algorithm is constrained by the hard capacity limit $|\mathcal{A}| \le K$, the selection probabilities must satisfy $\sum_{i \in [n]} x'_i \le K$. In addition, the probability $x'_i$ lies in $[0,1]$. Therefore, $\mathbf{x}' \triangleq [x'_i]_{i \in [n]}$ is a feasible solution to Problem (\ref{eq:fluid}), implying that $\texttt{OPT}(\mathcal{I}) \ge \texttt{LU}_{\mathcal{I}}(\mathbf{x}') = \texttt{OFF}(\mathcal{I})$.

Next, to show $\texttt{OPT}(\mathcal{I}) \le \texttt{OFF}(\mathcal{I})$, we note that the optimal solution $\mathbf{x}^*$ to Problem (\ref{eq:fluid}) can be used to construct an offline algorithm with a performance of $\texttt{LU}_{\mathcal{I}}(\mathbf{x}^*)$. 
%Specifically, the offline algorithm simulates a policy that makes randomized selections. 
Since the offline algorithm has complete knowledge of $\mathcal{I}$, it can compute $\mathbf{x}^*$. The dependent-rounding scheme of \citet{gandhi2006dependent} then produces a random set $\mathcal{A}$ such that $|\mathcal{A}|\le K$ almost surely and $\Pr[i\in\mathcal{A}]=x_i^*$ for every $i\in[n]$. Hence, every expected dimension utility is preserved, and the resulting offline algorithm achieves
\(
    \texttt{LU}_{\mathcal{I}}(\mathbf{x}^*)=\texttt{OPT}(\mathcal{I}).
\)

Thus, we show that the performance of the optimal offline algorithm is equal to $\texttt{OPT}(\mathcal{I})$.
\hfill \Halmos

\subsection{Proof of Proposition~\ref{prop:general_impossible}.}
\label{subsec:proof_general_impossible}

Fix $d\ge2$ and error factors $\underline{e},\bar{e}\ge1$ with $\Delta=\underline{e}\bar{e}>1$. Choose an integer $M$ large enough that
\[
    \bar{e}M-\frac{M+1}{\underline{e}}\ge1 .
\]
Such an $M$ exists because $\Delta>1$. The interval
\(
    [(M+1)/\underline{e},\,\bar{e}M]
\)
therefore contains an integer. Fix any integer $N$ in this interval.
Set $K=1$ and $c_k=1$ for every $k\in[d]$. For each hidden dimension $m\in[d]$, we define an instance $\mathcal{I}_m$ as follows. The common prefix consists of $M$ singleton candidates of type $\mathbf{e}_1$, then $M$ singleton candidates of type $\mathbf{e}_2$, and so on through $M$ singleton candidates of type $\mathbf{e}_d$. After this common prefix, instance $\mathcal{I}_m$ has one final candidate of type $\mathbf{1}-\mathbf{e}_m$, i.e., the candidate has every attribute except attribute $m$.

The true marginal counts in $\mathcal{I}_m$ are
\[
    \phi_m([n])=M,\qquad
    \phi_k([n])=M+1,\quad k\ne m.
\]
We give the policy the same prediction vector in every instance, namely
\[
    \widehat{\phi}_k=N,\qquad k\in[d].
\]
This prediction is valid for every hidden instance. Indeed, if $k=m$, then
\[
    \frac{\phi_k([n])}{\underline{e}}
    =\frac{M}{\underline{e}}
    \le N
    =\widehat{\phi}_k
    \le\bar{e}M
    =\bar{e}\phi_k([n]).
\]
If $k\ne m$, then
\[
    \frac{\phi_k([n])}{\underline{e}}
    =\frac{M+1}{\underline{e}}
    \le N
    =\widehat{\phi}_k
    \le \bar{e}(M+1)
    =\bar{e}\phi_k([n]).
\]
Hence the same prediction vector satisfies the required multiplicative error guarantee for all instances $\{\mathcal{I}_m\}_{m\in[d]}$.

We next upper bound the performance of any policy. Because the common prefix and the prediction vector are identical across all hidden instances, the policy behaves identically during the prefix. Let $P_h$ be the expected total selection probability assigned to the $M$ singleton candidates of type $\mathbf{e}_h$ during this common prefix. The hard capacity constraint $K=1$ implies
\[
    \sum_{h=1}^d P_h\le1.
\]
Therefore, there exists a dimension $m^*\in[d]$ such that $P_{m^*}\le1/d$. Consider instance $\mathcal{I}_{m^*}$. Attribute $m^*$ appears only in the prefix singleton candidates of type $\mathbf{e}_{m^*}$. The final candidate has zero utility in this dimension. Thus,
\[
    \texttt{ALG}(\mathcal{I}_{m^*})\le P_{m^*}\le \frac{1}{d}.
\]

On the other hand, the offline fractional solution that assigns probability $1/2$ to one singleton candidate of type $\mathbf{e}_{m^*}$ and probability $1/2$ to the final candidate of type $\mathbf{1}-\mathbf{e}_{m^*}$ is feasible because $K=1$. Every dimension receives utility at least $1/2$, and hence
\[
    \texttt{OPT}(\mathcal{I}_{m^*})\ge \frac{1}{2}.
\]
Consequently,
\[
    \frac{\texttt{ALG}(\mathcal{I}_{m^*})}{\texttt{OPT}(\mathcal{I}_{m^*})}
    \le \frac{1/d}{1/2}
    =\frac{2}{d}.
\]
This proves the claimed $O(1/d)$ upper bound under any nonzero multiplicative error in the marginal advice.
\hfill \Halmos

\subsection{Proof of Theorem~\ref{thm:impossible}}
\label{subsec:proof_speical_imossible}

It suffices to prove the claim for perfect-square values of $d$. Let $q=\sqrt{d}$ and partition the dimensions into $q$ disjoint groups $\mathcal{G}_1,\dots,\mathcal{G}_q$, each with $q$ dimensions. Set $K=1$ and $c_k=1$ for every $k\in[d]$. For any set of dimensions $\mathcal{X}\subseteq[d]$, let $\mathbf{t}_{\mathcal{X}}$ denote the type with ones exactly in dimensions in $\mathcal{X}$.

We construct $q$ instances, indexed by $m\in[q]$. The first $q$ candidates are identical across all instances. Candidate $h$ has type $\mathbf{t}_{\mathcal{G}_h}$ for each $h\in[q]$. Instance $\mathcal{I}_m$ then contains $q$ unit candidates, one for each dimension in $\mathcal{G}_m$, followed by one candidate of type $\mathbf{t}_{[d]\setminus \mathcal{G}_m}$. Thus, in every instance and for every dimension $k$, exactly two candidates possess attribute $k$. Consequently, the aggregate advice is identical across all instances, with $\phi_k([n])=2$ for all $k\in[d]$.

We first lower bound the offline benchmark. In instance $\mathcal{I}_m$, the fractional solution that assigns probability $1/2$ to the first candidate of type $\mathbf{t}_{\mathcal{G}_m}$ and probability $1/2$ to the final candidate of type $\mathbf{t}_{[d]\setminus \mathcal{G}_m}$ is feasible because $K=1$. Every dimension receives utility at least $1/2$, and hence
\begin{align}
\label{eq:sqrt_lower_bound_opt}
    \texttt{OPT}(\mathcal{I}_m)\ge \frac{1}{2},\qquad \forall m\in[q].
\end{align}

Now fix any policy. Since the first $q$ candidates and the aggregate advice are identical across the $q$ instances, the policy's behavior on these candidates is also identical across instances. Let $p_h$ be the probability that the policy selects the $h$-th candidate in this common prefix. The hard capacity $K=1$ implies $\sum_{h=1}^q p_h\le 1$, so there exists an index $m^*\in[q]$ such that $p_{m^*}\le 1/q$.

Consider instance $\mathcal{I}_{m^*}$. For dimensions in $\mathcal{G}_{m^*}$, utility can come only from the prefix candidate of type $\mathbf{t}_{\mathcal{G}_{m^*}}$ and from the corresponding unit candidates that arrive later. The final candidate has no attributes in $\mathcal{G}_{m^*}$. Let $r_k$ be the probability that the policy selects the unit candidate for dimension $k\in\mathcal{G}_{m^*}$. Again by the hard capacity constraint, $\sum_{k\in\mathcal{G}_{m^*}} r_k\le 1$. Therefore,
\begin{align*}
    \sum_{k\in\mathcal{G}_{m^*}} \mathbb{E}[\phi_k(\mathcal{A})]
    = q p_{m^*}+\sum_{k\in\mathcal{G}_{m^*}} r_k
    \le 2.
\end{align*}
Hence, for some dimension $k^*\in\mathcal{G}_{m^*}$,
\begin{align*}
    \mathbb{E}[\phi_{k^*}(\mathcal{A})]\le \frac{2}{q}.
\end{align*}
By the definition of $\texttt{ALG}$ as the minimum expected utility over dimensions, it follows that $\texttt{ALG}(\mathcal{I}_{m^*})\le 2/q$. Combining this inequality with \eqref{eq:sqrt_lower_bound_opt}, we obtain
\begin{align*}
    \frac{\texttt{ALG}(\mathcal{I}_{m^*})}{\texttt{OPT}(\mathcal{I}_{m^*})}
    \le \frac{2/q}{1/2}
    = \frac{4}{\sqrt{d}}.
\end{align*}
This proves the claimed $O(1/\sqrt{d})$ upper bound.
\hfill \Halmos

\section{Omitted Details from Section~\ref{sec:general}}
\subsection{Proof of Lemma~\ref{lemma:opt_bound}}
\label{subsec:proof_opt_bound}
For any feasible $\mathbf{x}$ and $k\in[d]$,
\[
    c_k\sum_{i\in[n]}t_{ik}x_i
    \le c_k\min\{\phi_k([n]),K\},
\]
which gives $\texttt{OPT}(\mathcal I)\le\overline{\texttt{OPT}}$. Also, $\min\{K,\phi_k([n])\}\le d\min\{K/d,\phi_k([n])\}$ for every $k$, and hence $\overline{\texttt{OPT}}\le d\,\underline{\texttt{OPT}}$. If $\underline{\texttt{OPT}}=0$, the result follows immediately. Otherwise, define
\[
    x'_i\triangleq
    \max_{k\in[d]}\left\{t_{ik}\min\left\{1,\frac{K/d}{\phi_k([n])}\right\}\right\}.
\]
Then $\mathbf{x}'\in[0,1]^n$ and, for every $k\in[d]$,
\begin{align*}
    \sum_{i\in[n]}x'_i
    &\le\sum_{k\in[d]}\frac{K/d}{\phi_k([n])}\sum_{i\in[n]}t_{ik}=K,\\
    c_k\sum_{i\in[n]}t_{ik}x'_i
    &\ge c_k\phi_k([n])\min\left\{1,\frac{K/d}{\phi_k([n])}\right\}
    =c_k\min\{K/d,\phi_k([n])\}.
\end{align*}
Thus $\mathbf{x}'$ is feasible and $\underline{\texttt{OPT}}\le\texttt{OPT}(\mathcal I)$.

The bound $\overline{\texttt{OPT}}\le d\,\underline{\texttt{OPT}}$ implies $R\le\max\{1,\lceil\log_2d\rceil\}$. Let $r^*$ be the largest index such that $\gamma^{(r^*)}\le\texttt{OPT}(\mathcal I)$. If $r^*<R$, then $2\gamma^{(r^*)}=\gamma^{(r^*+1)}>\texttt{OPT}(\mathcal I)$. Otherwise, $\texttt{OPT}(\mathcal I)\le\overline{\texttt{OPT}} \le2^R\underline{\texttt{OPT}}=2\gamma^{(R)}$.
\hfill \Halmos

\let\savedNonbinaryHsection\theHsection
\let\savedNonbinaryHsubsection\theHsubsection
\let\savedNonbinaryHequation\theHequation
\let\savedNonbinaryHlemma\theHlemma
\renewcommand{\theHsection}{nonbinary.\arabic{section}}
\renewcommand{\theHsubsection}{nonbinary.\arabic{section}.\arabic{subsection}}
\renewcommand{\theHequation}{nonbinary.\arabic{equation}}
\renewcommand{\theHlemma}{nonbinary.\arabic{lemma}}

\subsection{Nonbinary Attributes}
\label{app:nonbinary}

This section proves Theorems~\ref{thm:nonbinary_positive} and~\ref{thm:nonbinary_impossibility}. For the analysis, define
\[
    a_{ik}=t_{ik}/\ell,\qquad
    A_k=\sum_{i\in[n]}a_{ik}=S_k/\ell,\qquad
    \bar c_k=\ell c_k.
\]
Thus, \(a_{ik}\in\{0\}\cup[1,\kappa]\), where \(\kappa=u/\ell\), and this change of units preserves every dimension utility, as shown by
\[
    \bar c_k\sum_i a_{ik}x_i
    =c_k\sum_i t_{ik}x_i.
\]
Define
\[
    \underline{\texttt{OPT}}
    =\min_{k\in[d]}\bar c_k\min\{A_k,K/d\},
    \qquad
    \overline{\texttt{OPT}}
    =\min_{k\in[d]}\bar c_k\min\{A_k,\kappa K\}.
\]
If \(A_k=0\) for some \(k\), then $\texttt{OPT}(\mathcal{I}) = 0$ and thus Theorem~\ref{thm:nonbinary_positive} holds trivially. Hence, we restrict to the case where \(A_k>0\) for every \(k\).
% \subsubsection{Advice-Computable Bounds.}
Similar to Lemma~\ref{lemma:opt_bound}, it is easy to show that
\begin{lemma}
\label{lemma:nonbinary_opt_bounds}
We have
\(
    \underline{\emph{\texttt{OPT}}}
    \le \emph{\texttt{OPT}}(\mathcal I)
    \le \overline{\emph{\texttt{OPT}}}
    \le d\kappa\,\underline{\emph{\texttt{OPT}}}.
\)
\end{lemma}

% \proof{Proof of Lemma~\ref{lemma:nonbinary_opt_bounds}.}
% Let \(q_0\triangleq K/d\) and define
% \[
%     \widetilde x_i
%     \triangleq
%     \max_{k\in[d]}
%     \min\left\{1,\frac{q_0a_{ik}}{A_k}\right\}.
% \]
% Since the maximum is at most the sum,
% \[
%     \sum_i\widetilde x_i
%     \le
%     \sum_{k\in[d]}\sum_i
%     \min\left\{1,\frac{q_0a_{ik}}{A_k}\right\}
%     \le
%     \sum_{k\in[d]}\frac{q_0}{A_k}\sum_i a_{ik}
%     =K.
% \]
% Moreover, every positive \(a_{ik}\) is at least one. Hence, for every
% dimension \(k\),
% \begin{align*}
%     \sum_i a_{ik}\widetilde x_i
%     &\ge
%     \sum_{i:a_{ik}>0}
%     a_{ik}\min\left\{1,\frac{q_0a_{ik}}{A_k}\right\}\\
%     &\ge
%     \sum_{i:a_{ik}>0}
%     a_{ik}\min\left\{1,\frac{q_0}{A_k}\right\}
%     =\min\{A_k,q_0\}.
% \end{align*}
% Thus, \(\widetilde{\mathbf x}\) is a feasible witness for
% \(\underline{\texttt{OPT}}\).

% For any feasible \(\mathbf x\),
% \[
%     \sum_i a_{ik}x_i\le A_k,
%     \qquad
%     \sum_i a_{ik}x_i
%     \le\kappa\sum_i x_i
%     \le\kappa K,
% \]
% which proves the upper bound. Finally,
% \[
%     \min\{A_k,\kappa K\}
%     \le d\kappa\min\{A_k,K/d\}
% \]
% for every \(k\). Applying this inequality to a dimension attaining
% \(\underline{\texttt{OPT}}\) proves the last claim.
% \hfill \Halmos

\subsubsection{Positive Guarantee.}

We use exactly the principal-agent construction in Section~\ref{population-fractional-algorithm} and record only the objects that change. For agent \(r\), let
\[
    v_{ik}^{(r)}
    \triangleq
    \bar c_k\sum_{j<i}y_j^{(r)}a_{jk}.
\]
The active set and its weighted score are
\begin{align}
\label{eq:nonbinary_active_score}
    \mathcal Y_i^{(r)}(\tau)
    &\triangleq
    \left\{k\in[d]:
    a_{ik}>0,
    v_{ik}^{(r)}+\bar c_ka_{ik}\tau<\theta_r
    \right\},&
    G_i^{(r)}(\tau)
    &\triangleq
    \sum_{k\in\mathcal Y_i^{(r)}(\tau)}a_{ik}.
\end{align}
Controlled greedy uses \(G_i^{(r)}\) in place of active-set cardinality. For minimalist protection, define
\begin{align}
\label{eq:nonbinary_protection}
    w_{ik}^{(r)}
    &\triangleq
    \bar c_k\left(
    \sum_{j\le i}y_j^{(r)}a_{jk}
    +\sum_{j<i}z_{jk}^{(r)}a_{jk}
    \right),\notag\\
    \texttt{Res}_{ik}
    &\triangleq
    \bar c_k\left(A_k-\sum_{j\le i}a_{jk}\right),\notag\\
    m_{ik}^{(r)}(\tau)
    &\triangleq
    w_{ik}^{(r)}
    +\bar c_ka_{ik}\tau
    +\texttt{Res}_{ik}.
\end{align}
If \(a_{ik}=0\), set \(z_{ik}^{(r)}=0\). Otherwise, increase the supplemental selection fraction \(z_{ik}^{(r)}\in[0,1]\) until it reaches one, the protection budget is exhausted, or \(m_{ik}^{(r)}(z_{ik}^{(r)})\ge\theta_r\).
% As in the binary policy, the protection budget charges \(z_{ik}^{(r)}\), whereas its weighted contribution is \(z_{ik}^{(r)}a_{ik}\). 
The agent combination, the capped sum, and online rounding are unchanged.

For completeness, we record the weighted policy's parameters. When \(\underline{\texttt{OPT}}>0\), let
\[
    R\triangleq
    \max\left\{1,
    \left\lceil
    \log_2
    \frac{\overline{\texttt{OPT}}}
         {\underline{\texttt{OPT}}}
    \right\rceil\right\},
    \qquad
    \gamma^{(r)}
    \triangleq
    2^{r-1}\underline{\texttt{OPT}},
    \quad r\in[R],
\]
and, as in \eqref{eq:definition_weight}, set
\[
    \rho_r\triangleq
    \sqrt{\frac{\gamma^{(r)}}
                {\underline{\texttt{OPT}}}},
    \qquad
    W\triangleq\sum_{r=1}^R\rho_r,
    \qquad
    B_r\triangleq\frac{\rho_rK}{W},
    \qquad
    \theta_r\triangleq\frac{\gamma^{(r)}}{W}.
\]
The usual geometric-grid argument gives an \(r^*\in[R]\) such that
\[
    \frac{\texttt{OPT}(\mathcal I)}{2}
    \le\gamma^{(r^*)}
    \le\texttt{OPT}(\mathcal I).
\]
Lemma~\ref{lemma:nonbinary_opt_bounds} also implies \(2^R\le2d\kappa\), and therefore
\begin{align}
\label{eq:nonbinary_W}
    W
    =\frac{2^{R/2}-1}{\sqrt2-1}
    \le(2+\sqrt2)\sqrt{d\kappa}.
\end{align}

\proof{Proof of Theorem~\ref{thm:nonbinary_positive}.}
For agent \(r\), the greedy and protection budgets give
\[
    \sum_i x_i^{(r)}
    \le
    \frac12\left(
    \sum_i y_i^{(r)}
    +\sum_i\sum_kz_{ik}^{(r)}
    \right)
    \le B_r.
\]
Since \(\sum_rB_r=K\), the capped sum is feasible and dominates every agent componentwise. It remains to analyze the successful agent \(r^*\). Suppress its superscript and write \(\rho,B,\theta,\gamma\) for its parameters.

First suppose that controlled greedy exhausts its budget. Define
\[
    D_k
    \triangleq
    \sum_i\int_0^{y_i}
    a_{ik}\mathbbm{1}
    [k\in\mathcal Y_i(\tau)]\,d\tau .
\]
The active condition in \eqref{eq:nonbinary_active_score} gives \(D_k\le\theta/\bar c_k\). Because \(\underline{\texttt{OPT}}\le\bar c_kK/d\) for every \(k\),
\[
    \sum_kD_k
    \le\sum_k\frac{\theta}{\bar c_k}
    \le\frac{\gamma K}
             {W\underline{\texttt{OPT}}}
    =\frac{\rho^2K}{W}
    =\rho B.
\]
On the other hand, the weighted active score is at least \(\rho\) whenever the greedy budget is spent. Therefore,
\[
    \sum_kD_k
    =\sum_i\int_0^{y_i}G_i(\tau)\,d\tau
    \ge\rho\sum_i y_i
    =\rho B.
\]
Equality holds throughout, so \(D_k=\theta/\bar c_k\) for every \(k\). Consequently,
\(
    \bar c_k\sum_i a_{ik}y_i\ge\theta
\)
for all dimensions, and the agent's final vector attains at least \(\theta/2\).

Now suppose that the greedy budget is not exhausted, and let
\(
    Z\triangleq\sum_{i,k}z_{ik}.
\)
If \(Z<B\), the protection budget never binds. The post-protection potential for dimension \(k\) after arrival \(i\) is
\[
    M_{ik}
    \triangleq
    \bar c_k\left(
    \sum_{j\le i}y_ja_{jk}
    +\sum_{j\le i}z_{jk}a_{jk}
    +\sum_{j>i}a_{jk}
    \right).
\]
Initially,
\(
    M_{0k}=\bar c_kA_k
    \ge\texttt{OPT}(\mathcal I)
    \ge\gamma\ge\theta.
\)
Before processing protection at arrival \(i\), the potential changes by \(\bar c_k(y_i-1)a_{ik}\). If this takes it below \(\theta\), increasing \(z_{ik}\) to at most one can replace the entire loss. Induction therefore gives \(M_{ik}\ge\theta\) after every arrival. At the end of the horizon,
\[
    \bar c_k\left(
    \sum_i y_ia_{ik}+\sum_i z_{ik}a_{ik}
    \right)\ge\theta,
\]
so the agent again attains at least \(\theta/2\).

It remains to consider \(Z=B\). Define the final weighted shortfall
\[
    s_k
    \triangleq
    \left(
    \frac{\theta}{\bar c_k}
    -\sum_i y_ia_{ik}
    \right)^+,
    \qquad
    \mathcal Q\triangleq\{k:s_k>0\}.
\]
Let \(\mathbf x^*\) be an optimal fractional solution. Since \(\gamma\le\texttt{OPT}(\mathcal I)\),
\[
    \frac{\theta}{\bar c_k}
    \le\frac1W\sum_i a_{ik}x_i^*,
    \qquad k\in[d].
\]
If \(y_i<x_i^*/W\), neither the candidate cap nor the greedy budget can have stopped the greedy process. Moreover, every supported dimension in \(\mathcal Q\) remains active when candidate \(i\)'s greedy decision ends. The weighted-score stopping condition therefore gives \(\sum_{k\in\mathcal Q}a_{ik}<\rho\), and hence
\begin{align}
\label{eq:nonbinary_shortfall}
    \sum_ks_k
    \le
    \sum_i\left(\frac{x_i^*}{W}-y_i\right)^+
    \sum_{k\in\mathcal Q}a_{ik}
    \le
    \frac{\rho}{W}\sum_i x_i^*
    \le\frac{\rho K}{W}
    =B.
\end{align}

The exact residual in \eqref{eq:nonbinary_protection} also ensures that protection never supplies more weighted mass than the final shortfall.
\begin{align}
\label{eq:nonbinary_no_oversupply}
    \sum_i a_{ik}z_{ik}\le s_k,\qquad k\in[d].
\end{align}
To see this, a dimension outside \(\mathcal Q\) never needs protection because its final greedy mass already reaches the target. For \(k\in\mathcal Q\), consider the first arrival at which its greedy mass to date plus all remaining attribute mass falls below the target. Before this arrival, no protection is made. At this arrival the post-protection potential is at most \(\theta\), and at each later arrival its value before protection changes by \(\bar c_k(y_i-1)a_{ik}\le0\). The protection rule stops at \(\theta\), or earlier if a cap binds, so the post-protection potential remains at most \(\theta\). Evaluating it at the horizon gives \eqref{eq:nonbinary_no_oversupply}.

Finally, \(a_{ik}\ge1\) whenever \(z_{ik}>0\). Combining \eqref{eq:nonbinary_shortfall} and \eqref{eq:nonbinary_no_oversupply} yields
\[
    B=Z
    =\sum_{i,k}z_{ik}
    \le\sum_{i,k}a_{ik}z_{ik}
    \le\sum_ks_k
    \le B.
\]
All inequalities are tight. Thus, \(\sum_i a_{ik}z_{ik}=s_k\) for every \(k\), every dimension reaches the auxiliary target \(\theta\), and the successful agent attains at least \(\theta/2\). In all cases,
\[
    \texttt{LU}_{\mathcal I}(\mathbf x^{(r^*)})
    \ge\frac{\gamma^{(r^*)}}{2W}
    \ge\frac{\texttt{OPT}(\mathcal I)}{4W}.
\]
The capped sum preserves this value. Combining \eqref{eq:nonbinary_W} with Lemma~\ref{lemma:prophet} proves
\[
    \texttt{ALG}(\mathcal I)
    \ge
    \frac{\texttt{OPT}(\mathcal I)}
         {4(2+\sqrt2)\sqrt{d\kappa}}.
\]
\hfill \Halmos

\subsubsection{Impossibility Result.}

\proof{Proof of Theorem~\ref{thm:nonbinary_impossibility}.}
We first give a parameterized construction. Fix positive integers \(g,h,q\) satisfying \(gh\le d\) and \(q\le\kappa\). Partition the first \(gh\) dimensions into \(g\) groups \(\mathcal G_1,\ldots,\mathcal G_g\), each of size \(h\), and call the remaining dimensions dummy dimensions. Set \(K=1\), \(c_k=1\) for all \(k\), and
\[
    b\triangleq\frac{u}{q}.
\]
Since \(q\le u/\ell\), we have \(b\in[\ell,u]\).

For each hidden group \(m\in[g]\), construct an instance \(\mathcal I_m\). The common prefix consists of \(g\) candidates \(P_1,\ldots,P_g\), where \(P_j\) has value \(u\) exactly on the dimensions in \(\mathcal G_j\). The suffix of \(\mathcal I_m\) contains \(q\) singleton candidates of value \(b\) for each dimension in \(\mathcal G_m\), followed by a complement candidate \(C_m\) of value \(u\) on every core dimension outside \(\mathcal G_m\) and on every dummy dimension. If dummy dimensions exist, the suffix also contains one candidate of value \(u\) on all dummy dimensions and zero elsewhere.

These instances have the same horizon length and the same weighted marginal advice. Indeed, a dimension in \(\mathcal G_m\) receives \(u\) from \(P_m\) and total mass \(qb=u\) from its singleton candidates. Any other core dimension receives \(u\) from its prefix candidate and \(u\) from \(C_m\). A dummy dimension receives \(u\) from \(C_m\) and \(u\) from the dummy candidate. Hence,
\[
    S_k=2u,\qquad k\in[d],
\]
in every hidden instance. The fractional solution assigning one half to \(P_m\) and one half to \(C_m\) is feasible and gives every dimension utility at least \(u/2\). Therefore,
\begin{align}
\label{eq:nonbinary_lower_opt}
    \texttt{OPT}(\mathcal I_m)\ge u/2.
\end{align}

Fix any randomized online policy. Its behavior on the common prefix is identical across all hidden instances. Let \(p_j\) be the marginal probability with which it selects \(P_j\). Since \(K=1\), \(\sum_jp_j\le1\), so some \(m^*\) satisfies \(p_{m^*}\le1/g\). This choice depends only on the policy's distribution, not on its realized random bits.
For each \(k\in\mathcal G_{m^*}\), let \(r_k\) be the expected total selection probability assigned to its \(q\) singleton candidates. Again, \(\sum_{k\in\mathcal G_{m^*}}r_k\le1\). Neither \(C_{m^*}\) nor the dummy candidate (when present) has value in \(\mathcal G_{m^*}\), and therefore
\begin{align*}
    \texttt{ALG}(\mathcal I_{m^*})
    \le
    \frac1h\sum_{k\in\mathcal G_{m^*}}
    \mathbb E[\phi_k(\mathcal A)]
    =up_{m^*}+\frac{b}{h}
    \sum_{k\in\mathcal G_{m^*}}r_k
    \le\frac{u}{g}+\frac{u}{qh}.
\end{align*}
Together with \eqref{eq:nonbinary_lower_opt}, this gives
\begin{align}
\label{eq:nonbinary_parameterized_lower}
    \frac{\texttt{ALG}(\mathcal I_{m^*})}
         {\texttt{OPT}(\mathcal I_{m^*})}
    \le\frac{2}{g}+\frac{2}{qh}.
\end{align}

It remains to choose the parameters. Let
\[
    q\triangleq\min\{\lfloor\kappa\rfloor,d\},\qquad
    h\triangleq\left\lfloor\sqrt{\frac dq}\right\rfloor,\qquad
    g\triangleq\left\lfloor\frac dh\right\rfloor.
\]
Then \(gh\le d\), \(h\ge\frac12\sqrt{d/q}\), and \(g\ge d/(2h)\). Substitution in \eqref{eq:nonbinary_parameterized_lower} yields
\[
    \frac{2}{g}+\frac{2}{qh}
    \le\frac{8}{\sqrt{dq}}.
\]
Finally, if \(\kappa<d\), then \(q=\lfloor\kappa\rfloor\ge\kappa/2\). If \(\kappa\ge d\), then \(q=d\).
Thus, in either case,
\(
    q\ge\frac12\min\{\kappa,d\}.
\)
For every online policy, some instance with the same exact advice satisfies
\[
    \frac{\texttt{ALG}(\mathcal I)}
         {\texttt{OPT}(\mathcal I)}
    \le
    \frac{8\sqrt2}
         {\sqrt{d\min\{\kappa,d\}}},
\]
which proves the claimed impossibility result.
\hfill \Halmos

\let\theHsection\savedNonbinaryHsection
\let\theHsubsection\savedNonbinaryHsubsection
\let\theHequation\savedNonbinaryHequation
\let\theHlemma\savedNonbinaryHlemma

\section{Extension: Nash Social Welfare}
\label{app:nash_extension}

This section presents the policy for the Nash social welfare objective. The policy follows that in \citep{banerjee2022proportionally} by replacing horizon-dependent set-aside allocations with a set-aside calibrated by the aggregate advice. 

For a feasible fractional solution $\mathbf{x}$, define its Nash social welfare by
\begin{align*}
    {\texttt{NSW}}_{\mathcal{I}}(\mathbf{x}) \triangleq \Bigl(\prod_{k=1}^d U_k(\mathbf{x})\Bigr)^{1/d}.
\end{align*}
Let
\begin{align}
\label{eq:nash_fluid}
\tag{Nash-Fluid}
{\texttt{OPT}}_{{\texttt{NSW}}}(\mathcal{I}) \triangleq
\max_{\mathbf{x}} \quad & {\texttt{NSW}}_{\mathcal{I}}(\mathbf{x)}\\
\text{s.t.} \quad & \sum_{i\in[n]} x_i \le K,\notag\\
& 0\le x_i\le 1,\quad \forall i\in[n].\notag
\end{align}
If $\phi_k([n])=0$ for some dimension $k$, then $\texttt{OPT}_{\texttt{NSW}}(\mathcal{I})=0$ and the guarantee holds trivially. Hence, throughout this section, we focus on the nontrivial case $\phi_k([n])>0$ for every $k\in[d]$.

\paragraph{Our policy.}

For each dimension $k$, define the advice-calibrated set-aside rate and the guaranteed base utility
\begin{align}
\label{eq:nash_base}
    a_k \triangleq \min\Bigl\{1,\frac{K}{d\phi_k([n])}\Bigr\},
    \qquad
    \delta_k \triangleq \frac{c_k}{2}\phi_k([n])a_k
    =\frac{c_k}{2}\min\Bigl\{\phi_k([n]),\frac{K}{d}\Bigr\}.
\end{align}
When candidate $i$ arrives, the policy first assigns the set-aside probability
\begin{align}
\label{eq:nash_y}
    y_i\triangleq \frac{1}{2}\max_{k:t_{ik}=1} a_k,
\end{align}
with the convention that $y_i=0$ if candidate $i$ has no attributes. 

The second component is a Nashian greedy component. Let
\begin{align*}
    P_{ik}(z)\triangleq \delta_k+c_k\sum_{j<i}z_j t_{jk}+c_kzt_{ik}
\end{align*}
be the promised utility of dimension $k$ if the policy assigns greedy probability $z$ to candidate $i$. Fix
\begin{align*}
    \alpha_N \triangleq 2\log(2d+2),\qquad \lambda\triangleq \frac{\alpha_N}{K}.
\end{align*}
Given the previous greedy probabilities, choose $z_i\in[0,1-y_i]$ by increasing $z$ from zero until either $z=1-y_i$ or
\begin{align}
\label{eq:nash_threshold}
    \frac{1}{d}\sum_{k=1}^d \frac{c_kt_{ik}}{P_{ik}(z)}\le \lambda.
\end{align}
Finally, set
\begin{align}
\label{eq:nash_x}
    x_i \triangleq y_i+z_i
\end{align}
and apply the online level-set rounding scheme from Lemma~\ref{lemma:prophet}. The rounding step preserves the marginal probabilities and therefore preserves the expected utilities used by the Nash objective.

\setcounter{repeattheorem}{\getrefnumber{thm:geometric_mean}}
\addtocounter{repeattheorem}{-1}
\begin{repeattheorem}
\label{thm:nash_extension}
For the advice-augmented scenario, the policy defined by \eqref{eq:nash_base}--\eqref{eq:nash_x} is feasible and satisfies
\begin{align*}
    \emph{\texttt{NSW}}_{\mathcal{I}}(\mathbf{x})
    \ge
    \frac{\emph{\texttt{OPT}}_{\emph{\texttt{NSW}}}(\mathcal{I})}
    {1+\log(2d+1)+2\log(2d+2)}
    \ge
    \frac{\emph{\texttt{OPT}}_{\emph{\texttt{NSW}}}(\mathcal{I})}{4\log(2d+2)}.
\end{align*}
Consequently, after online level-set rounding, the same guarantee holds for Nash social welfare evaluated on the expected dimension utilities of the randomly selected set.
\end{repeattheorem}

\proof{Proof of Theorem~\ref{thm:nash_extension}.}
We first prove feasibility. By replacing the maximum in \eqref{eq:nash_y} with a summation,
\begin{align*}
    \sum_{i\in[n]} y_i
    \le \frac{1}{2}\sum_{i\in[n]}\sum_{k\in[d]} t_{ik}a_k
    =\frac{1}{2}\sum_{k\in[d]}\phi_k([n])a_k
    \le \frac{K}{2}.
\end{align*}
It remains to show that the Nashian greedy component also consumes at most $K/2$. Let
\begin{align*}
    P_k \triangleq \delta_k+c_k\sum_{i\in[n]} z_it_{ik}
\end{align*}
be the final promised utility of dimension $k$, and define the potential
\begin{align*}
    \Psi\triangleq \frac{1}{d}\sum_{k=1}^d \log\frac{P_k}{\delta_k}.
\end{align*}
During any infinitesimal increase of a greedy probability, the derivative of $\Psi$ is exactly the left-hand side of \eqref{eq:nash_threshold}, and hence is at least $\lambda$. Therefore, if the greedy component consumed total probability $K/2$, we would have
\begin{align*}
    \Psi\ge \lambda K/2=\log(2d+2).
\end{align*}
On the other hand, as long as the total greedy probability is at most $K/2$, we have for every dimension $k$,
\begin{align}
\label{eq:nash_potential_upper}
    \frac{P_k}{\delta_k}
    \le 2d+1.
\end{align}
Indeed, if $\phi_k([n])\le K/d$, then $P_k\le \delta_k+c_k\phi_k([n])\le 3\delta_k\le (2d+1)\delta_k$. If $\phi_k([n])>K/d$, then $P_k\le \delta_k+c_kK/2=(d+1)\delta_k$. Thus $\Psi\le\log(2d+1)<\log(2d+2)$, a contradiction. Hence the greedy component never reaches total probability $K/2$, and the vector $\mathbf{x}$ is feasible.

Next, let $Y_k\triangleq c_k\sum_i y_it_{ik}$ and $Z_k\triangleq c_k\sum_i z_it_{ik}$. The final utility of dimension $k$ is $U_k(\mathbf{x})=Y_k+Z_k$, and the construction of the set-aside component ensures $Y_k\ge \delta_k$ for every $k$. Consider any feasible fractional solution $\mathbf{w}$, and define
\begin{align*}
    s_i\triangleq \frac{1}{d}\sum_{k=1}^d \frac{c_kt_{ik}}{U_k(\mathbf{x})}.
\end{align*}
Since the greedy budget is not exhausted, if $x_i<1$, then the stopping condition \eqref{eq:nash_threshold} and the inequality $U_k(\mathbf{x})\ge P_k$ imply $s_i\le \lambda$. If $x_i=1$, then no feasible solution can assign more probability to candidate $i$. Therefore, using $r_i\triangleq (w_i-x_i)^+$, we get
\begin{align}
\label{eq:nash_ratio_bound}
\frac{1}{d}\sum_{k=1}^d\frac{U_k(\mathbf{w})}{U_k(\mathbf{x})}
= \sum_{i\in[n]} w_is_i 
\le \sum_{i\in[n]} x_is_i+\lambda\sum_{i\in[n]} r_i 
\le \sum_{i\in[n]} y_is_i+\sum_{i\in[n]} z_is_i+\lambda K .
\end{align}
The first term in the last line is at most one because
\begin{align*}
    \sum_{i\in[n]} y_is_i
    = \frac{1}{d}\sum_{k=1}^d \frac{Y_k}{U_k(\mathbf{x})}
    \le 1.
\end{align*}
For the second term, using $U_k(\mathbf{x})\ge P_k$ and the inequality $1-1/a\le \log a$ for $a\ge1$, we have
\begin{align*}
    \sum_{i\in[n]} z_is_i
    =\frac{1}{d}\sum_{k=1}^d \frac{Z_k}{U_k(\mathbf{x})}
    \le \frac{1}{d}\sum_{k=1}^d \frac{P_k-\delta_k}{P_k}
    \le \frac{1}{d}\sum_{k=1}^d \log\frac{P_k}{\delta_k}
    =\Psi
    \le \log(2d+1),
\end{align*}
where the last inequality follows from \eqref{eq:nash_potential_upper}. Substituting these bounds into \eqref{eq:nash_ratio_bound} gives
\begin{align}
\label{eq:nash_pf_bound}
    \frac{1}{d}\sum_{k=1}^d\frac{U_k(\mathbf{w})}{U_k(\mathbf{x})}
    \le 1+\log(2d+1)+2\log(2d+2).
\end{align}
Finally, by the AM-GM inequality,
\begin{align*}
    \frac{\texttt{NSW}_{\mathcal{I}}(\mathbf{w})}{\texttt{NSW}_{\mathcal{I}}(\mathbf{x})}
    =
    \Bigl(\prod_{k=1}^d\frac{U_k(\mathbf{w})}{U_k(\mathbf{x})}\Bigr)^{1/d}
    \le
    \frac{1}{d}\sum_{k=1}^d\frac{U_k(\mathbf{w})}{U_k(\mathbf{x})}.
\end{align*}
Taking $\mathbf{w}$ to be an optimal solution to \eqref{eq:nash_fluid} proves the first bound. Since $\log(2d+2)\ge1$ and $\log(2d+1)\le\log(2d+2)$ for $d\ge1$, the denominator is at most $4\log(2d+2)$. Lemma~\ref{lemma:prophet} then transfers the fractional guarantee to randomized online selections because it preserves all marginal selection probabilities.
\hfill \Halmos

\section{Extension: Order-Optimal Guarantee for Negative Generalized Means}
\label{app:nash_to_harmonic}

This section gives the complete policy and proof of Theorem~\ref{thm:negative_generalized_mean}. We retain the principal-agent construction and notation introduced in Section~\ref{subsubsec:negative_mean_policy}, and focus on the details needed to implement and analyze the policy.

\paragraph{Notation.}
Fix $q>0$ and write $p=-q$. Recall from Section~\ref{subsubsec:negative_mean_policy} that, for every nonempty $D\subseteq[d]$, $M_{-q}^{D}(\mathbf{x})$ is the negative mean over the dimensions in $D$, $\texttt{OPT}_{-q}^{D}(\mathcal I)$ is its offline benchmark over the original fractional feasible region, and
\(
    T_D\triangleq\texttt{OPT}_{-q}^{D}(\mathcal I).
\)
Let $T\triangleq T_{[d]}$, let $\mathbf{x}^*$ attain $T$, and set $U_k^*\triangleq U_k(\mathbf{x}^*)$. We consider the nontrivial case $T>0$. Then
\begin{align}
\label{eq:rp_identity}
    \frac1d\sum_{k=1}^d\Bigl(\frac{T}{U_k^*}\Bigr)^q=1 ,
\end{align}
and the analogous identity holds for $T_D$ on every reference set $D$. Recall the utility upper bound \(V_k\)
\(
    V_k\triangleq c_k\min\{\phi_k([n]),K\}
\)
which is computable from the marginal counts and upper-bounds $U_k(\mathbf{x})$ for every feasible $\mathbf{x}$. Hence
\begin{align}
\label{eq:rp_capacity}
    \frac1d\sum_{k=1}^d\Bigl(\frac{T}{V_k}\Bigr)^q\le1 .
\end{align}
We use
\begin{align*}
    \xi\triangleq\frac{q}{q+1},
    \qquad
    \rho\triangleq\frac{q}{2q+1},
    \qquad
    \frac{q\xi}{1+\xi}=q\rho .
\end{align*}
\subsection{The policy}
\label{app:nh_policy}

\paragraph{Grids.}
For every nonempty $D\subseteq[d]$, define the bounds computable from the marginal counts
\begin{align*}
    \underline T_D
    &\triangleq
    \left(
    \frac1{|D|}\sum_{k\in D}
    \left[c_k\min\left\{\phi_k([n]),\frac{K}{|D|}\right\}\right]^{-q}
    \right)^{-1/q},\\
    \overline T_D
    &\triangleq
    \left(
    \frac1{|D|}\sum_{k\in D}V_k^{-q}
    \right)^{-1/q}.
\end{align*}
It is easy to see 
\(
    \underline T_D\le T_D\le\overline T_D\le |D|\underline T_D.
\)
Therefore, the grid
\[
    \mathcal Q_D
    \triangleq
    \left\{2^a\underline T_D:
    a=0,\ldots,\left\lceil\log_2|D|\right\rceil\right\}
\]
contains a value $\Gamma_D$ satisfying $T_D/2\le\Gamma_D\le T_D$.

% \paragraph{Agents.}
Fix $\lambda\triangleq2^{-1/q}$ and $J\triangleq\lceil\log_2d\rceil+1$. For each global guess $\Gamma\in\mathcal Q_{[d]}$, form the capacity groups
\begin{align}
\label{eq:rp_groups}
    G_{\mathrm{hi}}(\Gamma)&\triangleq\{k:V_k\ge\lambda\Gamma\},\notag\\
    G_\ell(\Gamma)&\triangleq
    \{k:\lambda^{\ell+1}\Gamma\le V_k<\lambda^\ell\Gamma\},
    \qquad \ell=1,\ldots,J-1.
\end{align}
Empty groups are omitted. As in Section~\ref{subsubsec:negative_mean_policy}, $G$ denotes the dimensions protected by an agent, whereas $D$ denotes the reference set for its Nashian stage. For the high-capacity group, use
\[
    G=G_{\mathrm{hi}}(\Gamma),
    \qquad D=[d],
    \qquad \Gamma_D=\Gamma.
\]
For each lower-capacity group, use $G=D=G_\ell(\Gamma)$ and create one agent for every reference guess $\Gamma_D\in\mathcal Q_D$. Thus, an agent is determined by a global guess, one of its induced capacity groups, and a reference-optimum guess. The total number of agents is
\[
    P=O(\log^3d).
\]

\paragraph{The agent's algorithm.}
Give every agent a Nashian budget and a protection budget, each equal to
\[
    B\triangleq\frac{K}{2P}.
\]
The two stages operate online at every arrival. When candidate $i$ arrives, the agent first computes the Nashian allocation $x_i^N$ on its reference set $D$, using the algorithm of Appendix~\ref{app:nash_extension} with $(d,K)$ in its allocation parameters replaced by $(|D|,B)$. Let $\mathbf{x}^N$ denote the resulting Nashian vector and $U_k^N\triangleq U_k(\mathbf{x}^N)$. Define
\[
    L_{D,B}
    \triangleq
    1+\log(2|D|+1)+\frac{2K}{B}\log(2|D|+2),
    \qquad
    L\triangleq\max_D L_{D,B}
    =\mathrm{polylog}(d),
\]
where the maximum is over the agents' reference sets. Let
\[
    \beta
    \triangleq
    a_q\left(\frac{\lambda}{PdL^\xi}\right)^{1/(1+\xi)},
\]
where $a_q>0$ is a sufficiently small constant depending only on $q$. Any additional $q$-dependent constant in the protection target is absorbed into $a_q$. We set the protection target to
\[
    \tau\triangleq\beta\Gamma_D,
\]
for every dimension in $G$.

After computing $x_i^N$, the agent protects the dimensions in $G$ in a fixed order. For a trial supplemental allocation $s\in[0,(1-x_i^N)t_{ik}]$ to dimension $k$, define
\[
    m_{ik}^A(s)
    \triangleq
    c_k\sum_{j\le i}x_j^Nt_{jk}
    +c_k\sum_{j<i}z_{jk}^At_{jk}
    +c_kst_{ik}
    +c_k\sum_{j>i}t_{jk}.
\]
This is the utility already supplied by the Nashian and protection stages, plus the maximum additional utility available from candidates yet to arrive, as computed from the known marginal total. The agent starts from $z_{ik}^A=0$ and increases it continuously until either
\[
    z_{ik}^A=(1-x_i^N)t_{ik}
    \qquad\text{or}\qquad
    m_{ik}^A(z_{ik}^A)\ge\tau
\]
holds, or until any further increase would violate the protection-budget constraint
\[
    \sum_{j\le i}\sum_{k'\in G}z_{jk'}^A\le B.
\]
The agent then outputs
\[
    x_i^A\triangleq x_i^N+\max_{k\in G}z_{ik}^A.
\]
% Thus, unlike the max-min construction in Section~\ref{subsubsec:determine_z}, the Nashian allocation is retained in full rather than averaged with the protection allocation. Protection uses only the residual headroom of candidate $i$, and its budget is charged by the actual additional candidate mass $\max_{k\in G}z_{ik}^A$, rather than by the sum of the dimension-wise supplemental allocations.

\paragraph{The principal's allocation.}
The principal combines all agent vectors by the capped sum
\[
    x_i\triangleq\min\left\{1,\sum_Ax_i^A\right\}.
\]
It then applies online level-set rounding to $\mathbf{x}$.

\subsection{Proof of Theorem~\ref{thm:negative_generalized_mean}}
\label{app:nh_proof}

\proof{Proof.}
We first provide some preliminaries and then follow the three steps in the proof sketch in Section~\ref{subsubsec:negative_mean_policy}.

% \paragraph{Benchmark guesses.}
% Every feasible vector satisfies $U_k(\mathbf{x})\le V_k$, so $T_D\le\overline T_D$. The construction used in Lemma~\ref{lemma:opt_bound}, restricted to $D$, gives a feasible vector whose utility in dimension $k\in D$ is at least
% \(
%     c_k\min\{\phi_k([n]),K/|D|\}.
% \)
% Thus, $\underline T_D\le T_D$. Moreover,
% \[
%     |D|\min\left\{\phi_k([n]),\frac{K}{|D|}\right\}
%     =
%     \min\{|D|\phi_k([n]),K\}
%     \ge
%     \min\{\phi_k([n]),K\}
%     =
%     \frac{V_k}{c_k}.
% \]
% Hence, multiplying every utility in the vector defining $\underline T_D$ by $|D|$ dominates the vector defining $\overline T_D$, which gives
% \(
%     \overline T_D\le|D|\underline T_D.
% \)
% The claimed geometric grid and the bound $P=O(\log^3d)$ follow.

\paragraph{Reduced-budget Nashian certificate.}
For a Nashian run on $D$, let $\bar d=|D|$. The proof of Theorem~\ref{thm:nash_extension}, with $(d,K)$ in the algorithm replaced by $(\bar d,B)$, gives for every feasible comparator $\mathbf{w}$
\begin{align}
\label{eq:rp_nash_cert}
    \frac1{\bar d}\sum_{k\in D}
    \frac{U_k(\mathbf{w})}{U_k^N}
    \le
    L_{D,B}.
\end{align}
Indeed, the set-aside and greedy components each use at most $B/2$. In the performance analysis, only the last term on the right-hand side of \eqref{eq:nash_ratio_bound} changes. It is at most
\(
    2\log(2\bar d+2)K/B.
\)
Thus, \eqref{eq:rp_nash_cert} holds.

\paragraph{Step 1: Identify the dimensions that require protection.}
Fix a successful global guess $T/2\le\Gamma\le T$. Its groups cover $[d]$. Otherwise, some dimension would satisfy
\(
    V_k<\lambda^J\Gamma\le\lambda^JT,
\)
and hence $(T/V_k)^q>2^J>d$, contradicting \eqref{eq:rp_capacity}. For every nonempty lower group $G_\ell(\Gamma)$, set $D=G_\ell(\Gamma)$, write
\(
    T_\ell\triangleq T_D
    =\texttt{OPT}_{-q}^{G_\ell(\Gamma)}(\mathcal I)
\)
and fix a successful reference guess $T_D/2\le\Gamma_D\le T_D$. For the high group, set $D=[d]$ and $\Gamma_D=\Gamma$.

Consider one of these successful agents and suppress its superscript $A$. Let $\mathbf{x}^{*,D}$ attain $\texttt{OPT}_{-q}^{D}(\mathcal I)$ and set $U_k^{*,D}\triangleq U_k(\mathbf{x}^{*,D})$. Applying \eqref{eq:rp_nash_cert} to this solution and then applying H\"older's inequality with exponents $(q+1)/q$ and $q+1$ yields
\begin{align}
\label{eq:rp_moment}
    \frac1{|D|}\sum_{k\in D}
    \left(\frac{T_D}{U_k^N}\right)^\xi
    &\le
    \left(
    \frac1{|D|}\sum_{k\in D}
    \frac{U_k^{*,D}}{U_k^N}
    \right)^\xi
    \left(
    \frac1{|D|}\sum_{k\in D}
    \left(\frac{T_D}{U_k^{*,D}}\right)^q
    \right)^{1/(q+1)}
    \le L^\xi.
\end{align}
The second factor equals one by the definition of $T_D$. Since $\Gamma_D\le T_D$, every dimension with $U_k^N<\tau=\beta\Gamma_D$ is contained in
\(
    \{k:U_k^N<\beta T_D\}.
\)
Equation~\eqref{eq:rp_moment} therefore shows that at most
\begin{align}
\label{eq:rp_deficient}
    |D|(\beta L)^\xi
\end{align}
dimensions in $G$ require protection.

\paragraph{Step 2: Bound the protection cost.}
For the successful high-group agent,
\[
    \Gamma_D=\Gamma\le\frac{V_k}{\lambda},
    \qquad k\in G_{\mathrm{hi}}(\Gamma).
\]
For a successful lower-group agent $G=D=G_\ell(\Gamma)$,
\[
    \Gamma_D
    \le T_D
    \le\max_{j\in D}V_j
    <\lambda^\ell\Gamma
    \le\frac{V_k}{\lambda},
    \qquad k\in G_\ell(\Gamma).
\]
After decreasing $a_q$ if necessary, $\beta/\lambda\le1$. Hence $\tau\le V_k\le c_k\phi_k([n])$, so the target is feasible. For a deficient dimension $k$, the minimalist process stops adding supplemental allocation once its shortfall is covered (or when the protection budget binds). Therefore, its protection cost is at most
\[
    \sum_i z_{ik}
    \le\frac{\tau-U_k^N}{c_k}
    \le\frac{\tau}{c_k}
    \le
    \frac{\beta V_k}{\lambda c_k}
    \le
    \frac{\beta K}{\lambda}.
\]
Since $\max_{k\in G}z_{ik}\le\sum_{k\in G}z_{ik}$, combining this bound with \eqref{eq:rp_deficient} shows that the successful agent's total protection spend is at most
\[
    O_q\left(
    \frac{|D|}{\lambda}\beta^{1+\xi}L^\xi K
    \right)
    \le
    O_q\left(a_q^{1+\xi}\frac KP\right)
    <
    \frac{K}{2P}=B,
\]
where the last inequality follows by choosing $a_q$ sufficiently small. Thus, the protection budget of every successful agent remains unexhausted, and every dimension in its protected set reaches the target $\tau$.

% Every agent, successful or not, uses at most $B$ in its Nashian stage and at most $B$ in its protection stage. Therefore,
% \[
%     \sum_i x_i
%     \le
%     \sum_A\sum_i x_i^A
%     \le
%     P(2B)
%     =K,
% \]
% and the capped sum also ensures $0\le x_i\le1$. The principal's fractional allocation is feasible.

\paragraph{Step 3: Aggregate the group-level guarantees.}
The capped sum dominates each successful agent componentwise. Consequently, for every $k$ in that agent's protected set,
\[
    U_k(\mathbf{x})
    \ge U_k^N,
    \qquad
    U_k(\mathbf{x})
    \ge\beta\Gamma_D
    \ge\frac{\beta T_D}{2}.
\]
Absorbing the factor $1/2$ into a constant depending only on $q$ and using
\(
    \max\{a,b\}\ge a^{1-\xi}b^\xi
\)
gives
\[
    \left(\frac{T_D}{U_k(\mathbf{x})}\right)^q
    \le
    O_q\left(
    \beta^{-q\xi}
    \left(\frac{T_D}{U_k^N}\right)^\xi
    \right),
\]
where $q(1-\xi)=\xi$. Summing over $G\subseteq D$ and applying \eqref{eq:rp_moment} gives
\begin{align}
\label{eq:rp_local}
    \frac1{|D|}\sum_{k\in G}
    \left(\frac{T_D}{U_k(\mathbf{x})}\right)^q
    \le
    \Psi,
    \qquad
    \Psi\triangleq O_q\left(L^\xi\beta^{-q\xi}\right).
\end{align}

For the high-capacity group, $D=[d]$ and $T_D=T$, so \eqref{eq:rp_local} directly bounds its contribution to the global inverse-power sum by $\Psi$. For a lower-capacity group,
\[
    \frac1d\sum_{k\in G_\ell}
    \left(\frac{T}{U_k(\mathbf{x})}\right)^q
    \le
    \Psi\frac{|G_\ell|}{d}
    \left(\frac{T}{T_\ell}\right)^q.
\]
Because $\mathbf{x}^*$ belongs to the fractional feasible region used by $\texttt{OPT}_{-q}^{G_\ell}(\mathcal I)$,
\[
    \frac{|G_\ell|}{d}
    \left(\frac{T}{T_\ell}\right)^q
    \le
    \frac1d\sum_{k\in G_\ell}
    \left(\frac{T}{U_k^*}\right)^q.
\]
The lower groups are disjoint, so \eqref{eq:rp_identity} implies that these weights sum to at most one. Adding the high-group contribution gives
\[
    \frac1d\sum_{k=1}^d
    \left(\frac{T}{U_k(\mathbf{x})}\right)^q
    \le2\Psi.
\]
Finally,
\[
    \Psi
    =
    O_q\left(
    L^\xi
    \left(\frac{PdL^\xi}{\lambda}\right)^{q\xi/(1+\xi)}
    \right)
    =
    d^{q\rho}\mathrm{polylog}(d),
\]
because $P$ and $L$ are both polylogarithmic in $d$ and $q\xi/(1+\xi)=q\rho$. Therefore,
\[
    \frac{M_{-q}(\mathbf{x})}{T}
    =
    \left[
    \frac1d\sum_{k=1}^d
    \left(\frac{T}{U_k(\mathbf{x})}\right)^q
    \right]^{-1/q}
    \ge
    \frac{\Omega_q(1)}{d^\rho\mathrm{polylog}(d)}.
\]
Online level-set rounding preserves every expected dimension utility and hence the same negative-mean guarantee. This proves Theorem~\ref{thm:negative_generalized_mean}.
\hfill \Halmos

\subsection{Hardness for Negative Generalized Mean Objectives}
\label{app:pmean_hardness}

% This section proves the lower bound for finite negative generalized means. The logarithmic impossibility result for $0\le p\le1$ is given in Lemma~\ref{lem:positive_generalized_mean_lower}, due to \citet[Theorem~1]{banerjee2022proportionally}. At the egalitarian endpoint, Theorem~\ref{thm:impossible} gives the matching $O(d^{-1/2})$ obstruction.

% \setcounter{repeattheorem}{\getrefnumber{thm:negative_generalized_mean_lower}}
% \addtocounter{repeattheorem}{-1}
% \begin{repeattheorem}[\sc Polynomial Hardness for Negative Means]
% \label{prop:pmean_negative_hardness}
% Fix $q>0$ and write $p=-q$. There is a family of advice-augmented instances with binary attributes, unit capacity, and identical aggregate advice such that every algorithm has competitive ratio at most
% \begin{align*}
%     O_q\left(d^{-\frac{q}{2q+1}}\right).
% \end{align*}
% \end{repeattheorem}

\proof{Proof of Theorem~\ref{thm:negative_generalized_mean_lower}.}
Fix an arbitrary integer $d \ge 1$, and set
\begin{align*}
    g&\triangleq\left\lfloor d^{(q+1)/(2q+1)}\right\rfloor,
    &h&\triangleq\left\lfloor\frac{d}{g}\right\rfloor,
    &b&\triangleq\left\lfloor g^{1/(q+1)}\right\rfloor.
\end{align*}
Let $d_0\triangleq gh$. Partition the first $d_0$ dimensions into $g$ disjoint groups $\mathcal G_1,\dots,\mathcal G_g$, each of size $h$, and let $\mathcal R\triangleq[d]\setminus[d_0]$ be the remaining dimensions. For each subset $B\subseteq[g]$ with $|B|=b$, define
\begin{align*}
    H_B\triangleq\bigcup_{a\in B}\mathcal G_a .
\end{align*}
Instance $\mathcal I_B$ has two parts. The common prefix contains $g$ candidates $P_1,\dots,P_g$, where candidate $P_a$ has attributes exactly $\mathcal G_a$. The suffix contains one singleton candidate for every dimension $k\in H_B$, followed by one complement candidate $C_B$ with attributes $[d]\setminus H_B$. The capacity is one, and all utility coefficients are one.

All instances in the family have the same aggregate advice. Every core dimension $k\in[d_0]$ appears twice. One occurrence is in its prefix group candidate, and the other is either in its singleton candidate if $k\in H_B$ or in $C_B$ otherwise. Every remaining dimension $k\in\mathcal R$ appears once, in $C_B$. Thus, the advice vector is independent of $B$.

We first lower bound the offline benchmark. For a fixed $B$, consider the feasible fractional solution
\begin{align*}
    x(C_B)=\frac12,\qquad
    x(P_a)=\frac{1}{2b}\quad(a\in B),
\end{align*}
with all other candidates assigned probability zero. This solution uses total capacity one. Each dimension in $H_B$ receives utility $1/(2b)$, while each dimension outside $H_B$ receives utility $1/2$. Hence
\begin{align}
\label{eq:nh_hard_opt_power}
    \texttt{OPT}_{-q}(\mathcal I_B)^{-q}
    \le
    \frac{1}{d}\left(bh(2b)^q+(d-bh)2^q\right)
    =
    2^q\left(\frac{b^{q+1}h}{d}+1-\frac{bh}{d}\right).
\end{align}
Because $b^{q+1}\le g$ and $gh=d_0\le d$, the right-hand side is at most $2^{q+1}$. Therefore,
\begin{align}
\label{eq:nh_hard_opt}
    \texttt{OPT}_{-q}(\mathcal I_B)=\Omega_q(1).
\end{align}

Now fix an arbitrary algorithm. For each prefix group $a\in[g]$, let $\pi_a$ be the probability that the algorithm selects prefix candidate $P_a$ in the common prefix. These probabilities are the same across all instances because the prefix and the aggregate advice are identical. The hard capacity constraint implies $\sum_{a=1}^g\pi_a\le1$. Averaging over all $b$-subsets of $[g]$, there exists $B\subseteq[g]$ with $|B|=b$ such that
\begin{align}
\label{eq:nh_hard_prefix_average}
    \sum_{a\in B}\pi_a\le\frac{b}{g}.
\end{align}
We choose this $B$ as the adversarial instance.

Let $u_k$ be the expected utility of dimension $k$ under the algorithm on $\mathcal I_B$, and let $F\triangleq |H_B|=bh$. Dimensions in $H_B$ receive utility only from prefix candidates $P_a$ with $a\in B$ and from their singleton candidates. The complement candidate contributes nothing to $H_B$. Therefore,
\begin{align}
\label{eq:nh_hard_bad_sum}
    S_H\triangleq\sum_{k\in H_B}u_k
    \le
    h\sum_{a\in B}\pi_a+1
    \le
    \frac{hb}{g}+1,
\end{align}
where the term $1$ upper bounds the total expected selection probability spent on singleton candidates.

The algorithm's generalized-mean value on $\mathcal I_B$ is
\(
    \bigl(d^{-1}\sum_{k=1}^d u_k^{-q}\bigr)^{-1/q}.
\)
If any dimension has zero expected utility, this value is zero, and the result is immediate. Otherwise, Jensen's inequality for the convex function $u\mapsto u^{-q}$ gives
\begin{align}
\label{eq:nh_hard_alg_jensen}
    \frac{1}{d}\sum_{k=1}^d u_k^{-q}
    \ge
    \frac{1}{d}\sum_{k\in H_B}u_k^{-q}
    \ge
    \frac{F}{d}\left(\frac{F}{S_H}\right)^q .
\end{align}
Taking the $(-1/q)$-th power and using $F=bh$ and \eqref{eq:nh_hard_bad_sum}, we obtain
\begin{align}
\label{eq:nh_hard_alg_bound}
    \left(\frac{1}{d}\sum_{k=1}^d u_k^{-q}\right)^{-1/q}
    \le
    \left(\frac{d}{bh}\right)^{1/q}
    \left(\frac{1}{g}+\frac{1}{hb}\right).
\end{align}
The chosen parameters satisfy
\[
    b=\Theta_q\left(g^{1/(q+1)}\right),
    \qquad
    h=\Theta_q\left(g^{q/(q+1)}\right),
    \qquad
    d_0=gh=\Theta(d).
\]
Consequently, the right-hand side of \eqref{eq:nh_hard_alg_bound} is
\begin{align*}
    \left(\frac{d}{bh}\right)^{1/q}
    \left(\frac1g+\frac1{hb}\right)
    &=
    O_q\left(\left(\frac gb\right)^{1/q}\right)
    O_q\left(\frac1g\right)\\
    &=
    O_q\left(g^{1/(q+1)}\right)
    O_q\left(\frac1g\right)
    =
    O_q\left(g^{-q/(q+1)}\right)
    =
    O_q\left(d^{-q/(2q+1)}\right).
\end{align*}
Combining this bound with \eqref{eq:nh_hard_opt} proves that the algorithm's competitive ratio on $\mathcal I_B$ is $O_q(d^{-q/(2q+1)})$. Because the algorithm and $d$ were arbitrary, the theorem follows.
\hfill \Halmos

\end{document}